\documentclass[showpacs,twocolumn,amssymb,floatfix,pre,notitlepage]{revtex4-1}

\usepackage{amssymb,amsfonts,amsmath,bm}
\usepackage{graphics,graphicx}
\usepackage{sidecap}

\begin{document}

\title{Effective charge of cylindrical and spherical colloids immersed
in an electrolyte: the quasi-planar limit}

\author{Ladislav \v{S}amaj}
\altaffiliation[On leave from ]
{Institute of Physics, Slovak Academy of Sciences, Bratislava, Slovakia}
\author{Emmanuel Trizac}
\affiliation{Laboratoire de Physique Th\'eorique et Mod\`eles Statistiques,
UMR CNRS 8626, Universit\'e Paris-Sud, 91405 Orsay, France}

\begin{abstract}
We consider the non-linear Poisson-Boltzmann theory for a single cylindrical or spherical macro-ion
in symmetric 1:1, together with asymmetric 1:2 and 2:1 electrolytes. We focus on the regime
where $\kappa a $, the ratio of the macro-ion radius $a$ over the inverse Debye length in the bulk 
electrolyte, is large. 
Analyzing the structure of the analytical expansion emerging from a multiple scale analysis,
we uncover a hidden structure for the electrostatic potential. This structure, which appears after 
a heuristic resummation, suggests a new and convenient expansion scheme that we present and work out 
in detail. We show that novel exact results can thereby be obtained, in particular pertaining to 
effective charge properties, in complete agreement with the direct numerical solution
to the problem.
\end{abstract}


\date{\today}

\maketitle

\renewcommand{\theequation}{1.\arabic{equation}}
\setcounter{equation}{0}

\section{Introduction} \label{sec:Introduction}
Colloidal suspensions are composed of large and often highly charged macromolecules
(macro-ions or colloids), immersed in an electrolyte (``salt'') solution 
of mobile, positively or negatively charged micro-ions.
These micro-ions move in a solvent that is in a first approximation regarded as a 
medium of uniform dielectric permittivity. 
The system as a whole is assumed to be in thermal equilibrium at some inverse 
temperature $\beta = 1/(k_{\rm B}T)$.

In the 1920s, Debye and H\"uckel (DH) \cite{Debye} proposed a linearized
mean-field description of the bulk thermodynamics of Coulomb fluids
which is adequate in the high-temperature region $\beta\to 0$.
Some years earlier, Gouy \cite{Gouy10} and Chapman \cite{Chapman13} had established 
the nonlinear Poisson-Boltzmann (PB) mean-field treatment of the electric
double layer, which served as a basis for the DVLO theory of
colloidal interactions \cite{Verwey48}. 

A simple framework for studying the thermodynamics of colloidal suspensions
at finite density is provided by the cell model 
\cite{Belloni98, Hansen00, Lukatsky01, Levin02}.
If however the concentration of colloids in the system is very low, in the first 
approximation one can ignore their mutual interaction and consider 
the so-called infinite dilution limit.
Colloids can be then studied as isolated mesoscopic objects of a given shape
and bare charge, situated in a charged solution. This will be the viewpoint adopted here and
for simplicity, we will address the homogeneous dielectric case with 
the dielectric permittivity of the colloid $\epsilon'$ equal to that
of the solvent $\epsilon$ (no electrostatic image charges). 

The concept of (effective) charge renormalization, introduced by 
Alexander et al \cite{Alexander84} in the context of the PB cell model
is simple to define in the infinite dilution limit:
the renormalized/effective charge follows from the far-potential induced in the electrolyte
\cite{Diehl01,Trizac02,Bocquet02}. 
At large distances from a charged body with bare charge $Q_{\rm bare}$, 
the electrostatic potential takes the same form as
that obtained within the linearized DH theory, with a modified prefactor 
$Q_{\rm eff}$ which embodies the nonlinear effects of the PB theory, or of 
an approach which goes beyond the mean-field description.
Within the nonlinear PB approach, $Q_{\rm eff}\simeq Q_{\rm bare}$ for low
values of $|Q_{\rm bare}|$ while $Q_{\rm eff}$ saturates to a finite constant
when $|Q_{\rm bare}|\to\infty$.
In general, one expects that $\vert Q_{\rm eff}\vert \ll \vert Q_{\rm bare}\vert$
as a consequence of the nonlinear screening effect of the electric double
layer around a colloid.
For a monovalent 1:1 electrolyte the effective charge is indeed always smaller
than the bare one. 
This is no longer true for asymmetric electrolytes which exhibit an 
overshooting effect \cite{Tellez04}: there exists a (rather small) interval of 
$Q_{\rm bare}$ where $|Q_{\rm eff}| \ge | Q_{\rm bare}|$.
 
Although the definition of an effective charge is unambiguous within the 
nonlinear PB theory, it is not clear whether the far-potential behaves 
like the DH one in an exact description which goes beyond the mean-field
assumption. 
The two-dimensional symmetric Coulomb gas of pointlike $\pm q$ charges,
interacting via the logarithmic Coulomb potential, is integrable in the whole 
interval of couplings $0\le \Gamma\equiv \beta q^2 <2$ where oppositely
charged couples of charges do not collapse \cite{Samaj00}.
The concept of renormalized charge has been shown to be valid for a charged 
conductor wall \cite{Samaj05a}, a pointlike guest charge \cite{Samaj05b} 
and for a guest charge with a small hard core \cite{Tellez05,Samaj06} which
permits one to go beyond the stability threshold.
It is interesting that for a guest charge with a small hard core at 
a finite temperature \cite{Samaj06}, $Q_{\rm eff}$ turns out to 
be a non-monotonous function of $Q_{\rm bare}$; the same phenomenon was 
observed also in Monte-Carlo \cite{Groot91} and molecular-dynamics 
\cite{Diehl04} simulations of the salt-free cell model.
Moreover, as $Q_{\rm bare}$ goes to infinity, the effective charge does not
saturate to a value but oscillates between two extreme (minimal and maximal)
values.

In this paper, we restrict ourselves to the definition of the effective charge
within the nonlinear PB theory. 
For a charged infinite plane, one can obtain explicit results 
for the symmetric 1:1 and asymmetric 1:2 and 2:1 electrolytes \cite{Gouy10}.
For other asymmetric two-component electrolytes, the solution can be 
constructed implicitly, see a short review \cite{Tellez11}.
Realistic macro-ions are usually modeled as curved objects, namely 
cylinders or spheres of a given radius $a$.
For such systems, the functional relation between $Q_{\rm eff}$ and 
$Q_{\rm bare}$ depends, besides the salt content, also on the dimensionless
parameter $\kappa a$ where $\kappa$ is the inverse Debye (correlation) 
length of micro-particles.
Two limiting cases are studied:
\begin{itemize}
\item
If $\kappa a \ll 1$, the analysis is difficult due to the 
counter-ion evaporation phenomenon, that may in the no salt limit
be complete for spheres, or partial for cylinders,
see \cite{Ramanathan83} and \cite{Ramanathan88}. Yet,
the cylindrical PB equation is Painlev\'e integrable for the symmetric
1:1 and asymmetric 2:1 and 1:2 electrolytes.
This enables one to construct systematically the non-analytical
$\kappa a$-expansion of the effective charge \cite{Tellez06}.
\item
If $\kappa a \gg 1$, the colloid radius is large compared to the Debye
length $\kappa^{-1}$ and the plane geometry is a good reference for
finding $1/(\kappa a)$ expansions of the effective charge. 
Shkel et al \cite{Shkel00} constructed an asymptotic large-distance
expansion of cylindrical and spherical PB equations for the symmetric
1:1 electrolyte. 
Using the method of multiple scales, they were able to derive the first
two terms of the $1/(\kappa a)$ expansion of the electric potential. 
Based on this work, analytical approximations were developed in Ref.
\cite{Aubouy03} for the symmetric 1:1 electrolyte and in Ref. \cite{Tellez04}
for the asymmetric 2:1 and 1:2 electrolytes.
\end{itemize}

The present paper concentrates on the large $\kappa a$ limit, 
constructing expansions in $1/(\kappa a)$ 
for the effective charge of cylindrical and spherical colloids. 
The multiple-scale method applied in the previous Refs. 
\cite{Shkel00,Aubouy03,Tellez04} is laborious and in practice does not
allow to go to high orders of the $1/(\kappa a)$ expansions.
However, inspecting the structure of these results 
suggests some re-summation can be performed, which in turn strongly points to 
a novel re-parametrization of the electric potential.
Pushing further this idea, it appears that the algebra is conducive to a 
much easier derivation of the $1/(\kappa a)$ expansions.
For the cylindrical geometry especially, the formulation ends up 
with a representation of each expansion order in terms of finite polynomials,
which enables one to construct expansions to
arbitrary high orders.
As concerns the spherical geometry, we were able to go one order beyond
the results of \cite{Shkel00,Aubouy03,Tellez04}; as a by product 
of the analysis, a divergence problem
for higher-order terms indicates the change of the analytic behaviour
of the expansion. For both cylindrical and spherical geometries
and the three types of electrolytes (1:1, 1:2 and 2:1), 
the obtained analytical results for the coefficients of the 
$1/(\kappa a)$ series expansions are tested against numerical resolutions.

The article is organized as follows.
In Sec. \ref{Formalism}, we introduce basic formulae and definitions
which are used throughout the paper. It is elementary and may be skipped by the
reader familiar with the subject.
The large-distance formalism of Shkel et al \cite{Shkel00} and 
the ensuing possible re-parametrization 
of the PB potential, are then
explained in Sec. \ref{Expansion}.
Our approach is presented for 1:1 electrolytes in Sec. \ref{1:1},
for 2:1 electrolytes in Sec. \ref{2:1} and for 1:2 electrolytes in
Sec. \ref{1:2}.
A brief recapitulation and concluding remarks are given in Sec.
\ref{Conclusion}. 

\renewcommand{\theequation}{2.\arabic{equation}}
\setcounter{equation}{0}

\section{Basic formalism} \label{Formalism}

\subsection{Microscopic models}
In this paper, we shall consider the infinite dilution limit of colloids
in suspensions, namely a unique colloid immersed in an infinite electrolyte 
solution.
The system is formulated in the three-dimensional (3D) Euclidean space
\cite{rque50}.
Three colloidal shapes are of interest:
\begin{itemize}
\item 
The planar case, when the colloid occupies the half-space $z\le 0$.
The surface at $z=0$ bears fixed surface charge density
$\sigma e$, where $e$ denotes the elementary charge and $\sigma$ has
dimension $[{\rm length}]^{-2}$; without any loss of generality we assume
that $\sigma>0$.
\item 
The cylindrical geometry, when the colloid corresponds to an infinitely 
long cylinder of radius $a$, say along the $z$-axis, carrying a bare 
linear charge density $\lambda e$, $\lambda$ having dimension 
$[{\rm length}]^{-1}$.
The system has a polar symmetry in the $(x,y)$ plane.  
\item 
The spherical geometry where the colloid is a sphere of radius $a$
with center localized at the origin ${\bf 0}$ and carrying a bare charge
$Z e$. $Z$ is dimensionless
and the system is radially symmetric.  
\end{itemize}
The space outside the macro-ion is filled by an infinite electrolyte 
solution. 
In general, it consists of $M$ $(=2,3,\ldots)$ types of mobile 
(pointlike) micro-ions $\nu=1,2,\ldots,M$ with positive or negative 
charges $\{ q_{\nu} e \}$, where $\vert q_{\nu}\vert$ is the valence of 
$\nu$-species.
The charged particles are immersed in a solvent which is a medium of uniform
dielectric permittivity $\epsilon$ (in Gauss units, $\epsilon\simeq 80$
for water).
They interact with each other and with the charged colloid surface 
via the standard Coulomb potential 
$v({\bf r})=1/(\epsilon\vert {\bf r}\vert)$, which is the solution 
of the 3D Poisson equation
\begin{equation} \label{Poisson}
\Delta v({\bf r}) = - \frac{4\pi}{\epsilon} \delta({\bf r}) .
\end{equation}

The system is in thermal equilibrium at the inverse temperature 
$\beta=1/(kT)$; we denote by $\langle \cdots \rangle$ the statistical
averaging over a thermodynamic ensemble.
It is useful to introduce the so-called Bjerrum length
$l_{\rm B} \equiv \beta e^2/\epsilon$, i.e. the distance at which two unit 
charges interact with thermal energy $kT$.
The macroscopic (i.e. thermodynamically averaged over all possible
particle configurations) density of $\nu$-species at point ${\bf r}$ 
is defined by $n_{\nu}({\bf r}) = \langle \hat{n}_{\nu}({\bf r}) \rangle$, where 
$\hat{n}_{\nu}({\bf r}) = \sum_i \delta({\bf r}-{\bf r}_i) \delta_{\nu,\nu_i}$;
$i$ numerates the particles of species $\nu_i$ at spatial positions
${\bf r}_i$, $\delta$ denotes Dirac/Kronecker delta function/symbol and 
the hat in $\hat{n}$ means ``microscopic'' 
(i.e. for a given particle configuration). 
The charge density at point ${\bf r}$ is given by 
\begin{equation}
\rho({\bf r}) = \langle \hat{\rho}({\bf r}) \rangle , \qquad
\hat{\rho}({\bf r}) = \sum_{\nu} \hat{n}_{\nu}({\bf r}) q_{\nu} e .
\end{equation}
At large distances from an isolated colloid, i.e. in the bulk, the species
number densities become homogeneous, $n_{\nu}({\bf r})=n_{\nu}$, and
the requirement of the bulk electroneutrality is equivalent to
\begin{equation} \label{bulkneutral}
\rho({\bf r}) = \rho = \sum_{\nu} n_{\nu} q_{\nu} e = 0 .
\end{equation}

\subsection{Poisson-Boltzmann equation}
For a given charge density profile $\rho({\bf r}')$, the mean electrostatic 
potential $\psi$ at point ${\bf r}$ is given by
$\psi({\bf r}) = \int {\rm d}{\bf r}'\, v(\vert {\bf r}-{\bf r}'\vert)
\rho({\bf r}')$.
The potential satisfies a counterpart of 
the Poisson equation (\ref{Poisson}),
\begin{equation} \label{Poisson2}
\Delta \psi({\bf r})  = - \frac{4\pi}{\epsilon} \rho({\bf r}) .
\end{equation}
In the microscopic picture, the energy of an $\nu$-particle at point
${\bf r}$ can be expressed in terms of the microscopic potential
$\hat{\psi}({\bf r})$ as $q_{\nu} e \hat{\psi}({\bf r})$ and the
probability of finding the particle at ${\bf r}$ is proportional to
the Boltzmann factor $\exp[-\beta q_{\nu} e \hat{\psi}({\bf r})]$.
In a mean-field approach, one adopts this microscopic relation
to the corresponding macroscopic values,
$n_{\nu}({\bf r}) = n_{\nu} \exp[-\beta q_{\nu} e \psi({\bf r})]$;
the normalization by the bulk value is consistent with the assumption
that $\psi({\bf r})$ and its derivatives vanish in the bulk.
This relation is exact in the high-temperature limit and only approximative
for a finite temperature \cite{comment66}.
Considering it in the Poisson Eq. (\ref{Poisson2}), one obtains 
a self-consistent PB equation for the mean electrostatic potential:
\begin{equation}
\Delta \psi({\bf r}) = - \frac{4\pi e}{\epsilon} 
\sum_{\nu} n_{\nu} q_{\nu} \exp[-\beta q_{\nu} e \psi({\bf r})] .
\end{equation}
In terms of the reduced potential $\phi({\bf r}) \equiv \beta e \psi({\bf r})$
and the inverse Debye length
$\kappa = \sqrt{4\pi l_{\rm B} \sum_{\nu} n_{\nu} q_{\nu}^2}$,
the PB equation can be rewritten as 
\begin{equation} \label{PB}
\Delta \phi({\bf r}) = - \frac{\kappa^2}{\sum_{\nu} n_{\nu} q_{\nu}^2} 
\sum_{\nu} n_{\nu} q_{\nu} {\rm e}^{-q_{\nu} \phi({\bf r})} .
\end{equation}

All studied geometries are effectively one-dimensional problems.
Let $r$ be the distance from the plane in the planar case, the distance
$\sqrt{x^2+y^2}$ from the cylinder axis $z$ or the distance
$\sqrt{x^2+y^2+z^2}$ from the sphere center.
The Laplacian for such symmetric case can be written as
\begin{equation}
\Delta \to \frac{1}{r^{\alpha}}\frac{\rm d}{{\rm d}r} 
\left( r^{\alpha} \frac{\rm d}{{\rm d}r} \right) = \frac{{\rm d}^2}{{\rm d}r^2}
+ \frac{\alpha}{r} \frac{\rm d}{{\rm d}r} ,
\end{equation}
where $\alpha=0$ for the planar case, $\alpha=1$ for the cylindrical geometry 
and $\alpha=2$ for the spherical geometry.
The corresponding second-order differential equation (\ref{PB}) has to
be supplemented by two boundary conditions (BCs), one at the contact with
the colloid and the regularity one at an infinite distance from the colloid.
The best way to derive these BCs is to consider the overall electroneutrality
of the system.
\begin{itemize}
\item
{\bf The planar case:} Integrating the 1D Poisson equation
\begin{equation}
\frac{{\rm d}^2 \psi(r)}{{\rm d}r^2} = - \frac{4\pi}{\epsilon} \rho(r)
\end{equation}
over $r$ from $0$ to $\infty$, we get
\begin{equation}
\psi'(\infty) - \psi'(0) = - \frac{4\pi}{\epsilon} 
\int_0^{\infty} {\rm d}r\, \rho(r) .
\end{equation}
The requirement of the overall electroneutrality
\begin{equation}
\sigma e + \int_0^{\infty} {\rm d}r\, \rho(r) = 0 
\end{equation}
then implies the couple of BCs for the reduced potential
\begin{equation} \label{BC0}
\phi'(0) = - 4 \pi l_{\rm B} \sigma , \qquad 
\lim_{r\to\infty} \phi'(r) = 0 .
\end{equation}
\item
{\bf Cylindrical geometry:} For a given charge density profile of particles
at $r\ge a$, the electroneutrality condition reads as
\begin{equation}
\lambda e + \int_a^{\infty} {\rm d}r\, 2\pi r \rho(r) = 0 .
\end{equation}
Multiplying the 2D Poisson equation
\begin{equation}
\frac{1}{r}\frac{\rm d}{{\rm d}r} \left( r \frac{{\rm d}\psi}{{\rm d}r} 
\right) = - \frac{4\pi}{\epsilon} \rho(r)
\end{equation}
by $r$ and integrating over $r$ from $a$ to $\infty$, 
the condition of overall electroneutrality is consistent with two BCs
for the reduced potential
\begin{equation} \label{BC1}
a \phi'(a) = - 2 l_{\rm B} \lambda , \qquad \lim_{r\to\infty} r \phi'(r) = 0 . 
\end{equation}
\item
{\bf Spherical geometry:} The electroneutrality condition reads
\begin{equation}
Z e + \int_a^{\infty} {\rm d}r\, 4\pi r^2 \rho(r) = 0 .
\end{equation}
With regard to the 3D Poisson equation
\begin{equation}
\frac{1}{r^2}\frac{\rm d}{{\rm d}r} \left( r^2 \frac{{\rm d}\psi}{{\rm d}r} 
\right) = - \frac{4\pi}{\epsilon} \rho(r) ,
\end{equation}
the electroneutrality condition is equivalent to two BCs
\begin{equation} \label{BC2}
a^2 \phi'(a) = - l_{\rm B} Z , \qquad \lim_{r\to\infty} r^2 \phi'(r) = 0 . 
\end{equation}
\end{itemize}

\subsection{Effective charge}
The nonlinear PB equation (\ref{PB}) can be linearized by applying
the expansion $\exp[-q_{\nu}\phi({\bf r})]\sim 1 -q_{\nu}\phi({\bf r})$.
With regard to the bulk electroneutrality condition (\ref{bulkneutral}),
we arrive at the linear DH equation
\begin{equation} \label{DH}
\Delta \phi_{\rm DH}({\bf r}) = \kappa^2 \phi_{\rm DH}({\bf r}) , 
\end{equation}
whose form does not depend on the particular composition of the electrolyte.
This equation, supplemented by the appropriate boundary conditions,
is solvable explicitly for all considered geometries.

The linearization of the potential Boltzmann factor is not adequate
mainly in the neighbourhood of the colloid, where the potential is large.
On the other hand, at asymptotically large distances from the colloid 
the potential is infinitesimally small and the linearization procedure 
is exact.
We can say that the asymptotic PB solution satisfies the linear equation 
\begin{equation}
\Delta \phi({\bf r}) = \kappa^2 \phi({\bf r}) , \qquad 
\vert {\bf r}\vert \to \infty .
\end{equation}
Comparing with the DH equation (\ref{DH}) we see that the asymptotic forms
of the PB and DH solutions are equivalent, up to position-independent
prefactors:
\begin{equation}
\phi_{\rm DH}(r) \mathop{\sim}_{r\to\infty} A_{DH}(Q) f(\kappa r) , \qquad
\phi(r) \mathop{\sim}_{r\to\infty} A(Q) f(\kappa r) .
\end{equation}
Here, the dependence of the $A$-prefactors on the thermodynamic parameters
of the electrolyte like $\kappa a$ will not be explicitly indicated and $Q$ 
is the bare charge characteristics of the colloid (the surface charge 
density $\sigma e$ for the plane case $\alpha=0$, the line charge density 
$\lambda e$ for the cylinder $\alpha=1$ and the charge $Z e$ for 
the sphere $\alpha=2$).
 
The effective charge $Q_{\rm eff}$ is defined as a function of the bare charge
$Q$ via the formula
\begin{equation} \label{eff}
A(Q) = A_{\rm DH}(Q_{\rm eff}) .
\end{equation}
In other words, $Q_{\rm eff}(Q)$ is the effective charge in the linear DH
theory which reproduces the correct PB asymptotic behavior; $Q_{\rm eff}$  accounts for nonlinear effects,
most prevalent close to the surface of 
the colloid.
The nonlinear effects are negligible in the limit $Q\to 0$ and one expects that
$Q_{\rm eff}(Q) \mathop{\sim}_{Q\to 0} Q$.
In the opposite limit $Q\to\infty$ one anticipates that the saturation value
of the effective charge
\begin{equation}
Q_{\rm eff}^{\rm sat} \equiv \lim_{Q\to\infty} Q_{\rm eff}(Q)
\end{equation}
is finite.

Let us now assume that we know the prefactor $A(Q)$ and derive an explicit
formula for the effective charge for each of the three geometries.
\begin{itemize}
\item
{\bf The planar case:} The solution of the DH equation 
\begin{equation}
\phi''_{\rm DH}(r) = \kappa^2 \phi_{\rm DH}(r)
\end{equation}
with the BCs (\ref{BC0}) takes the form 
\begin{equation}
\phi_{\rm DH}(r) = \frac{4\pi l_{\rm B}\sigma}{\kappa}  {\rm e}^{-\kappa r} . 
\end{equation}
Let us choose $A_{\rm DH}(\sigma) = 4\pi l_{\rm B}\sigma/\kappa$ and
$f(\kappa r) ={\rm e}^{-\kappa r}$. 
Anticipating that the nonlinear PB potential behaves asymptotically as
\begin{equation} \label{asymPB}
\phi(r) \mathop{\sim}_{r\to\infty} A(\sigma) {\rm e}^{-\kappa r} ,
\end{equation}
using the prescription (\ref{eff}) the effective surface charge density
$\sigma_{\rm eff} e$ depends on the bare one $\sigma e$ as follows
\begin{equation} \label{sigmaeff}
\frac{4\pi l_{\rm B}}{\kappa} \sigma_{\rm eff} = A(\sigma) .
\end{equation}
\item
{\bf Cylindrical geometry:} The linearized DH equation
\begin{equation}
\phi''_{\rm DH}(r) + \frac{1}{r} \phi'_{\rm DH}(r) = \kappa^2 \phi_{\rm DH}(r)
\end{equation}
with the BCs (\ref{BC1}) provides the solution
\begin{equation} \label{DHcylinder}
\phi_{\rm DH}(r) = \frac{2\lambda l_{\rm B}}{\kappa a K_1(\kappa a)}
K_0(\kappa r) , \qquad r\ge a ,
\end{equation}
where $K_0$ and $K_1$ are the modified Bessel functions.
Since $K_0(\kappa r) \sim \sqrt{\pi/(2\kappa r)} {\rm e}^{-\kappa r}$ for
asymptotically large $r$ we can choose 
\begin{equation}
A_{\rm DH}(\lambda) = \frac{\sqrt{2\pi}}{\kappa a K_1(\kappa a)} 
\lambda l_{\rm B} , \qquad 
f(\kappa r) = \frac{1}{\sqrt{\kappa r}} {\rm e}^{-\kappa r} . 
\end{equation}
The prefactor to the asymptotic behaviour of the full PB potential
\begin{equation}
\phi(r) \mathop{\sim}_{r\to\infty} A(\lambda) 
\frac{1}{\sqrt{\kappa r}} {\rm e}^{-\kappa r} 
\end{equation}
determines the dependence of the effective line charge density
$\lambda_{\rm eff} e$ on the bare one $\lambda e$ as follows
\begin{equation} \label{eff1}
\lambda_{\rm eff} l_{\rm B} = \frac{1}{\sqrt{2\pi}} \kappa a K_1(\kappa a) 
A(\lambda) .
\end{equation} 
\item
{\bf Spherical geometry:} The linearized DH equation
\begin{equation}
\phi''_{\rm DH}(r) + \frac{2}{r} \phi'_{\rm DH}(r) = \kappa^2 \phi_{\rm DH}(r)
\end{equation}
with the BCs (\ref{BC2}) has the solution
\begin{equation} \label{DHsphere}
\phi_{\rm DH}(r) = \frac{Z \kappa l_{\rm B}}{1+\kappa a} 
\frac{1}{\kappa r} {\rm e}^{-\kappa (r-a)} , \qquad r\ge a .
\end{equation}
It is natural to choose
\begin{equation}
A_{\rm DH}(Z) = \frac{\kappa}{1+\kappa a} {\rm e}^{\kappa a} Z l_{\rm B} , \qquad 
f(\kappa r) = \frac{1}{\kappa r} {\rm e}^{-\kappa r} . 
\end{equation}
Taking into account the expected asymptotic behaviour of the PB potential 
\begin{equation}
\phi(r) \mathop{\sim}_{r\to\infty} A(Z) 
\frac{1}{\kappa r} {\rm e}^{-\kappa r}  ,
\end{equation}
the formula for the effective charge $Z_{\rm eff} e$ as the function of 
the bare charge $Z$ reads as
\begin{equation} \label{eff2}
Z_{\rm eff} \frac{l_{\rm B}}{a} = \frac{1+\kappa a}{\kappa a} 
{\rm e}^{-\kappa a} A(Z) .
\end{equation}
In other words, $Z_{\rm eff}$ is the value that should be plugged in
the right-hand side of \eqref{DHsphere}, so that the latter formula provides the correct 
far-field of the non-linear solution to the PB equation. By construction thus,
$Z_{\rm eff}=Z$ when the PB theory is linearizable, which is the case
for $Z\to 0$. 
\end{itemize}

\subsection{Explicit results for the planar case}
The planar case is solvable explicitly only for specific types of
two-component electrolytes.
\begin{itemize}
\item
{\bf Symmetric 1:1 electrolyte.} 
We have two types of particles with (reduced) positive $q_1=1$ and 
negative $q_2=-1$ charges.
Denoting by $n$ the total particle number density, the requirement of 
the bulk electroneutrality is equivalent to $n_1 = n_2 = n/2$. 
The inverse Debye length is given by $\kappa = \sqrt{4\pi l_{\rm B} n}$.
The corresponding PB equation 
\begin{equation}
\phi''(r) = \kappa^2 \sinh \phi(r)
\end{equation}
has the explicit solution
\begin{equation} \label{11potential}
\phi(r) = 2 \ln \left[ \frac{1+\frac{A(\sigma)}{4}{\rm e}^{-\kappa r}}
{1-\frac{A(\sigma)}{4}{\rm e}^{-\kappa r}} \right]
\end{equation}
which indeed behaves at large $\kappa r$ as predicted by 
formula (\ref{asymPB}).
The relation between $A$ and $\sigma$ is yielded by the BC (\ref{BC0}) at $r=0$:
\begin{equation}
A(\sigma) = \frac{8\pi l_{\rm B}\sigma}{\kappa}
\frac{1}{1+\sqrt{1+\left( \frac{2\pi l_{\rm B}\sigma}{\kappa} \right)^2}} .
\end{equation}
In accordance with the relation (\ref{sigmaeff}), the effective charge 
density is given by
\begin{equation}
\sigma_{\rm eff} = \frac{2\sigma}{1+\sqrt{1+\left( 
\frac{2\pi l_{\rm B}\sigma}{\kappa} \right)^2}} .
\end{equation}
It has the correct behaviour $\sigma_{\rm eff}\sim \sigma$ in the limit
$\sigma\to 0$.
In the saturation $\sigma\to\infty$ limit, we have
\begin{equation}
\frac{4\pi l_{\rm B}}{\kappa} \sigma_{\rm eff}^{\rm sat} = 4 .
\label{eq:sigmaeffsatplane}
\end{equation}
\item 
{\bf Asymmetric 2:1 electrolyte.}
In this case, the positively charged coions to the surface have $q_1=2$
and the negatively charged counterions have $q_2=-1$.
The bulk electroneutrality requires that $n_1=n/3$ and $n_2=2n/3$, where
$n$ is the total particle number density.  
The inverse Debye length $\kappa = \sqrt{8\pi l_{\rm B} n}$.
The PB equation 
\begin{equation}
\phi''(r) = \kappa^2 \frac{1}{3} \left[ 
{\rm e}^{\phi(r)}-{\rm e}^{-2\phi(r)} \right]
\end{equation}
has been solved by Gouy \cite{Gouy10}:
\begin{equation} \label{Gouy}
\phi(r) = \ln \left[ 1 + \frac{A(\sigma){\rm e}^{-\kappa r}}{\left(
1-\frac{A(\sigma)}{6}{\rm e}^{-\kappa r}\right)^2} \right] .
\end{equation}
The potential behaves at large $\kappa r$ as (\ref{asymPB}).
The relation between $A$ and $\sigma$ follows from the BC (\ref{BC0}) at $r=0$:
\begin{equation}
\frac{4\pi l_{\rm B}\sigma}{\kappa} = \frac{36 A (6+A)}{(6-A)(A^2+24 A +36)} .
\end{equation} 
From the three $A$-solutions of this cubic equation we take the one which
goes to zero in the limit $\sigma\to 0$.
In the saturation limit $\sigma\to\infty$, we have $A_{\rm sat}=6$ and
therefore 
\begin{equation}
\frac{4\pi l_{\rm B}}{\kappa} \sigma_{\rm eff}^{\rm sat} = 6 .
\end{equation}
\item 
{\bf Asymmetric 1:2 electrolyte.} 
Now the coions have $q_1=1$ and the bulk number density $n_1=2n/3$, while 
the counterions have $q_2=-2$ and $n_2=n/3$. 
As before, $\kappa = \sqrt{8\pi l_{\rm B} n}$.
The PB equation 
\begin{equation}
\phi''(r) = \kappa^2 \frac{1}{3} \left[ 
{\rm e}^{2\phi(r)}-{\rm e}^{-\phi(r)} \right]
\end{equation}
has the explicit solution
\begin{equation} \label{Gouy2}
\phi(r) = - \ln \left[ 1 - \frac{A(\sigma){\rm e}^{-\kappa r}}{\left(
1+\frac{A(\sigma)}{6}{\rm e}^{-\kappa r}\right)^2} \right] . 
\end{equation}
The relation between $A$ and $\sigma$, following from the BC (\ref{BC0}) 
at $r=0$, takes the form
\begin{equation}
\frac{4\pi l_{\rm B}\sigma}{\kappa} = \frac{36 A (6-A)}{(6+A)(A^2-24 A +36)} .
\end{equation} 
The physical $A$-solution goes to zero in the limit $\sigma\to 0$.
In the saturation limit $\sigma\to\infty$, $A_{\rm sat}=6(2-\sqrt{3})$ 
is the smaller root of the quadratic equation $A^2-24 A +36=0$ and we arrive at 
\begin{equation}
\frac{4\pi l_{\rm B}}{\kappa} \sigma_{\rm eff}^{\rm sat} = 6 (2-\sqrt{3}) .
\end{equation}
\end{itemize}

\renewcommand{\theequation}{3.\arabic{equation}}
\setcounter{equation}{0}

\section{An asymptotic expansion for 1:1 electrolyte} \label{Expansion}
For the symmetric 1:1 electrolyte of total particle number density $n$
and $\kappa=\sqrt{4\pi l_{\rm B}n}$, the general PB equation for 
the electrostatic potential $\phi$ reads as
\begin{equation} \label{generalPB}
\phi''(x) + \frac{\alpha}{x} \phi'(x) = \sinh \phi(x) ,
\end{equation}
where $x=\kappa r$ is the reduced distance, $\alpha=1$ for the cylindrical 
colloid and $\alpha=2$ for the spherical colloid.  
In the paper \cite{Shkel00}, an asymptotic large-$x$ expansion of the
potential has been obtained in the following form
\begin{eqnarray}
\phi(x) & = & \frac{{\rm e}^{-x}}{x^{\alpha/2}} \left(
A_{00} + \frac{A_{01}}{x} + \frac{A_{02}}{x^2} + \cdots \right) + \nonumber \\ 
& & + \frac{{\rm e}^{-3x}}{x^{3\alpha/2}} \left( A_{10} + \frac{A_{11}}{x} 
+ \frac{A_{12}}{x^2} + \cdots \right) + \cdots \label{asymptA} \\
& & + \frac{{\rm e}^{-(2j+1)x}}{x^{(2j+1)\alpha/2}} \left( A_{j0} + \frac{A_{j1}}{x} + 
\frac{A_{j2}}{x^2} + \cdots \right) + \cdots . \nonumber
\end{eqnarray}
Here, $A_{00}\equiv A$ is the crucial prefactor to the leading large-distance 
asymptotic.
The other prefactors $A_{jk}$ scale with $A$ like $A_{jk}= a_{jk} A^{2j+1}$, 
where the coefficients (numbers) $\{ a_{jk}\}$ fulfill a rather complicated 
recursion which enables one to generate systematically the coefficients
$\{ a_{jk}\}$ with small values of the $(j,k)$ indices \cite{Shkel00}. 

The first $j=0$ row of the asymptotic formula (\ref{asymptA}), which consists
in the exponential ${\rm e}^{-x}$ multiplied by an infinite inverse-power-law
series, corresponds to the linear DH approximation. 
The next rows are exponentially smaller and smaller corrections 
to the DH result.

For our purposes, it is more important to concentrate on columns.
Let us introduce the new variable
\begin{equation} \label{deft}
t(x) = \frac{A}{4} \frac{{\rm e}^{-x}}{x^{\alpha/2}}
\end{equation}
where $A$ (and thus $t$) implicitly depends on the bare charge 
($\lambda$ or $Z$ depending on the geometry), and $\kappa a$; we shall usually omit in 
the notation these functional dependences or write only the relevant ones.
In the BC, the first derivative of the potential is taken just at the surface
of the colloid, i.e. at $x=\kappa a$.
The value of $t$ is, in general, not small at the colloid surface, 
even in the limit of interest $\kappa a\to\infty$.
Note that the given column differs from the previous one basically by 
the factor $1/x$.
The corresponding surface factor $1/(\kappa a)$ is small in the limit 
$\kappa a\to\infty$. 
This permits one to generate a systematic expansion of the prefactor $A$ 
in $1/(\kappa a)$ powers by taking successively column by column.

The asymptotic expansion (\ref{asymptA}) can be rewritten in the variables
$t$ and $1/x$ as follows
\begin{equation} \label{potentialt}
\phi(x) = \sum_{j,k=0}^{\infty} t^{2j+1} B_{jk} \frac{1}{x^k} ,
\end{equation}
where the coefficients
\begin{equation} \label{asymptB}
B_{jk} \equiv \left( \frac{4}{A} \right)^{2j+1} A_{jk}
\end{equation}
are the numbers which will be explicitly available for small values
of $(j,k)$ indices.

We shall also need the asymptotic large-$x$ expansion of the modified
Bessel functions
\begin{equation} \label{Knu}
K_{\nu}(x) = \sqrt{\frac{\pi}{2x}} {\rm e}^{-x} \sum_{k=0}^{\infty} \frac{1}{(2x)^k}
\frac{\Gamma(\frac{1}{2}+\nu+k)}{k! \Gamma(\frac{1}{2}+\nu-k)} 
\end{equation}
for $\nu=0,1$.

\subsection{Cylindrical geometry}
For the $\alpha=1$ cylindrical geometry, we have the parameter
\begin{equation} \label{cylindert}
t(x) = \frac{A(\lambda)}{4} \frac{{\rm e}^{-x}}{\sqrt{x}} .
\end{equation}
The coefficients $B_{jk}$ with $j,k=0,1,2,3,4$ are summarized 
in Table \ref{Table1}. 
\begin{table}
\begin{tabular}{c||c|c|c|c|c} \hline \hline
\phantom{a} j \phantom{a} & k=0 & k=1 & k=2 & k=3 & k=4 
\\ \hline \hline & & & & & \\
0 & 4 & $-\dfrac{1}{2}$ & $\dfrac{9}{2^5}$ & $-\dfrac{75}{2^8}$ & 
$\dfrac{3675}{2^{13}}$ \\ 
& & & & & \\ \hline
& & & & & \\
1 & $\dfrac{4}{3}$ & $-\dfrac{3}{2}$ & $\dfrac{71}{2^5}$ & 
$-\dfrac{3215}{3\times 2^8}$ & $\dfrac{79 521}{2^{13}}$ \\ 
& & & & & \\ \hline
& & & & & \\ 
2 & $\dfrac{4}{5}$ & $-\dfrac{3}{2}$ & $\dfrac{89}{2^5}$ & $-\dfrac{1485}{2^8}$ 
& $\dfrac{114 071}{2^{13}}$ \\ 
& & & & & \\ \hline
& & & & & \\
3 & $\dfrac{4}{7}$ & $-\dfrac{3}{2}$ & $\dfrac{107}{2^5}$ & 
$-\dfrac{1965}{2^8}$ & $\dfrac{157 237}{2^{13}}$ \\ 
& & & & & \\ \hline
& & & & & \\
4 & $\dfrac{4}{9}$ & $-\dfrac{3}{2}$ & $\dfrac{125}{2^5}$ & 
$-\dfrac{2517}{2^8}$ & $\dfrac{211 275}{2^{13}}$ \\
& & & & & \\ \hline \hline
\end{tabular}
\caption{Cylindrical geometry. The coefficients $B_{jk}$ $(j,k=0,1,2,3,4)$ 
appearing in (\ref{asymptB}) for the large-distance expansion (\ref{asymptA}) 
obtained in Ref. \cite{Shkel00}.}
\label{Table1}
\end{table}
The first row of the table corresponds to the DH result (\ref{DHcylinder}).
Indeed, writing in (\ref{DHcylinder}) the large-$x$ expansion of the modified 
Bessel function (\ref{Knu}) with $\nu=0$, the corresponding coefficients 
are found to be
\begin{equation}
B_{0k} \equiv 4 \frac{A_{0k}}{A} = 2^{2-k}
\frac{\Gamma(\frac{1}{2}+k)}{k! \Gamma(\frac{1}{2}-k)},
\end{equation}
the first few of which read $4$, $-1/2$, $9/2^5$ etc.

In accordance with our strategy, let us consider the first column of 
Table \ref{Table1}.
From the first few coefficients $B_{j0}$ we can ``guess'' 
\begin{equation}
B_{j0} = \frac{4}{2j+1}
\end{equation}
and suggest that this formula holds for all $j=0,1,\ldots$.
If this is true, the potential (\ref{potentialt}) is given, in the leading
$1/x^0$ order, by
\begin{equation}
\phi(x) = 4 \sum_{j=0}^{\infty} \frac{t^{2j+1}}{2j+1} =
2 \ln \left( \frac{1+t}{1-t} \right) .
\end{equation}
This result is identical to the planar one (\ref{11potential}) under the only
proviso that the cylindrical $t$ (\ref{cylindert}) differs from the corresponding
planar function $A {\rm e}^{-x}/4$ by the factor $1/\sqrt{x}$.

We can go further and sum the contributions of the second column of
Table \ref{Table1} to determine the potential up to the $1/x$ order.
We anticipate that
\begin{equation}
B_{01} = - \frac{1}{2} , \qquad B_{j1} = - \frac{3}{2} \quad 
\mbox{for $j\ge 1$.}
\end{equation}
The potential (\ref{potentialt}) is then given by
\begin{eqnarray}
\phi(x) & = & 2\ln \left( \frac{1+t}{1-t} \right) -\frac{1}{2x} t
- \frac{3}{2x} \frac{t^3}{1-t^2} \nonumber \\ 
& = &  2\ln \left( \frac{1+t}{1-t} \right) -\frac{t(1+2t^2)}{4x} 
\left( \frac{1}{1-t} + \frac{1}{1+t} \right) . \phantom{aaa} \label{orig}
\end{eqnarray}
The $1/x$ correction consists of $1/(1-t)$ and $1/(1+t)$ terms.
They arise naturally from a ``renormalization ansatz'' 
\begin{equation} \label{finalphi1}
\phi(x) = 2\ln \left( \frac{1+f(x)}{1-f(x)} \right) 
\end{equation}
with
\begin{equation} \label{finalphi2}
f(x) = t(x) \left\{ 1 - \frac{1}{8x} [1 + 2 t^2(x)] \right\} .
\end{equation}
It is easy to verify that this ansatz coincides with the original 
equation (\ref{orig}) to the order $1/x$.

The guessing of the coefficients becomes harder when considering the next columns.
In the following section, we shall show how to generate systematically 
the whole infinite series of coefficients in a straightforward way.
At this stage, we restrict ourselves to the preliminary result 
(\ref{finalphi1}), (\ref{finalphi2}).

For the solution of type (\ref{finalphi1}), the BC (\ref{BC1}) at $r=a$ 
can be expressed as
\begin{equation} \label{Alambda}
-l_{\rm B} \lambda = 2 \kappa a \frac{f'(\kappa a)}{1-f^2(\kappa a)} .
\end{equation}
This relation determines the function $A(\lambda)$.
In the saturation limit $\lambda\to\infty$ we have $f_{\rm sat}(\kappa a)=1$,
as indicated by Eq. \eqref{Alambda}.
Consequently,
\begin{equation}
t_{\rm sat}(\kappa a)\left\{ 1 - \frac{1}{8\kappa a}
[1+2 t_{\rm sat}^2(\kappa a)] \right\} = 1 .
\end{equation} 
Performing the large-$\kappa a$ expansion in this formula, we obtain
\begin{equation} \label{tsat}
t_{\rm sat}(\kappa a) = 1 + \frac{3}{8 \kappa a} + \cdots 
\equiv \frac{A_{\rm sat}}{4} \frac{{\rm e}^{-\kappa a}}{\sqrt{\kappa a}} ,
\end{equation}
where $A_{\rm sat} \equiv A(\lambda\to\infty)$.
Using that
\begin{equation}
K_1(\kappa a) = \sqrt{\frac{\pi}{2\kappa a}} {\rm e}^{-\kappa a}
\left( 1 + \frac{3}{8\kappa a} + \ldots \right) ,
\end{equation}
the formula for the effective charge (\ref{eff1}) implies that its 
saturation value has the large-$\kappa a$ expansion of the form
\begin{eqnarray}
\lambda_{\rm eff}^{\rm sat} l_{\rm B} & = & \frac{1}{\sqrt{2\pi}} \kappa a 
K_1(\kappa a) A_{\rm sat} \nonumber \\ & = & 2\kappa a + \frac{3}{2} + \cdots .
\label{prescr1}
\end{eqnarray}
This result agrees with the previous finding of Ref. \cite{Aubouy03}.

\subsection{Spherical geometry}
For the $\alpha=2$ spherical geometry, we have
\begin{equation} \label{spheret}
t(x) = \frac{A}{4} \frac{{\rm e}^{-x}}{x} .
\end{equation}
The coefficients $B_{jk}$ with $j,k=0,1,2,3,4$ are summarized 
in Table \ref{Table2}. 

\begin{table}
\begin{tabular}{c||c|c|c|c|c} \hline \hline
\phantom{a} j \phantom{a} &  k=0 & k=1 & k=2 & k=3 & k=4 \\ \hline \hline
0 & 4 & 0 & 0 & 0 & 0 \\ \hline
& & & & & \\
1 & $\dfrac{4}{3}$ & $-2$ & $\dfrac{7}{2}$ & $-\dfrac{15}{2}$ & 
$\dfrac{155}{2^3}$ \\ 
& & & & & \\ \hline 
& & & & & \\
2 & $\dfrac{4}{5}$ & $-2$ & $\dfrac{14}{3}$ & $-\dfrac{35}{3}$ & 
$\dfrac{2327}{3^2\times 2^3}$ \\ 
& & & & & \\ \hline
& & & & & \\
3 & $\dfrac{4}{7}$ & $-2$ & $\dfrac{205}{3^2\times 2^2}$ & 
$-\dfrac{863}{3^3\times 2}$ & $\dfrac{40621}{3^3\times 2^5}$ \\ 
& & & & & \\ \hline 
& & & & & \\
4 & $\dfrac{4}{9}$ & $-2$ & $\dfrac{2413}{5\times 3^2\times 2^3}$ & 
$-\dfrac{90271}{5\times 3^3\times 2^5}$ & 
$\dfrac{11313029}{5^2\times 3^3\times 2^8}$ \\ 
& & & & & \\ \hline \hline
\end{tabular}
\caption{Spherical geometry. The coefficients $B_{jk}$ $(j,k=0,1,2,3,4)$
appearing in (\ref{asymptB}) for the large-distance expansion (\ref{asymptA}) 
obtained in Ref. \cite{Shkel00}.}
\label{Table2}
\end{table}

The first row of the table corresponds to the DH result (\ref{DHcylinder}),
namely $B_{0k} = 4 \delta_{0k}$.
The first column of Table \ref{Table2} is identical to the first one of
Table \ref{Table1}, so that
\begin{equation}
B_{j0} = \frac{4}{2j+1}
\end{equation}
and, in the leading $1/x^0$ order, the potential is given by
\begin{equation}
\phi(x) = 2 \ln \left( \frac{1+t}{1-t} \right) .
\end{equation}
Again, this result is identical to the planar one (\ref{11potential}) if the planar 
$t$ is multiplied by $1/x$.

It is easy to guess the second column of Table \ref{Table2}: 
\begin{equation}
B_{01} = 0 , \qquad B_{0j} = -2 \quad \mbox{for $j\ge 1$.}
\end{equation}
The potential (\ref{potentialt}) is then given by
\begin{equation}
\phi(x) =  2\ln \left( \frac{1+t}{1-t} \right) 
-\frac{t^3}{x} \left( \frac{1}{1-t}  + \frac{1}{1+t} \right) . 
\end{equation}
The potential is transformable to the renormalized form of 
type (\ref{finalphi1}) with 
\begin{equation} \label{f}
f(x) = t(x) \left[ 1 - \frac{1}{2x} t^2(x) \right] .
\end{equation}

As before, the saturation limit $Z\to\infty$ corresponds to 
$f_{\rm sat}(\kappa a)=1$.
Consequently,
\begin{equation}
t_{\rm sat}(\kappa a) \left[ 1 - \frac{1}{2\kappa a}
t_{\rm sat}^2(\kappa a) \right] = 1 .
\end{equation} 
The large-$\kappa a$ expansion of $t_{\rm sat}(\kappa a)$ then reads
\begin{equation}
t_{\rm sat}(\kappa a) = 1 + \frac{1}{2 \kappa a} + \cdots 
\equiv \frac{A_{\rm sat}}{4} \frac{{\rm e}^{-\kappa a}}{\kappa a} ,
\end{equation}
where $A_{\rm sat}\equiv A(Z\to\infty)$.
The formula for the effective charge (\ref{eff2}) then leads to
\begin{eqnarray}
Z_{\rm eff}^{\rm sat} \frac{l_{\rm B}}{a} & = & \frac{1+\kappa a}{\kappa a}
{\rm e}^{-\kappa a} A_{\rm sat} \nonumber \\
& = & 4\kappa a + 6 + \cdots \label{prescr2}
\end{eqnarray}
which coincides with the finding of Ref. \cite{Aubouy03}.
It is instructive to compute the corresponding effective surface charge 
$\sigma_{\rm eff}^{\rm sat}=Z_{\rm eff}^{\rm sat}/(4\pi a^2)$
and likewise for the cylinder,  
$\sigma_{\rm eff}^{\rm sat} = \lambda_{\rm eff}^{\rm sat}/(2\pi a)$.
In doing so, it appears that Eq. \eqref{prescr2} and Eq. \eqref{prescr1}
bear the same information.
Indeed, introducing the curvature ${\cal C}=1/a$ for cylinders and
${\cal C}=2/a$ for spheres, both can be written, when phrased in terms
of the effective surface charge, as
\begin{equation}
\sigma_{\rm eff}^{\rm sat} \,=\, \frac{\kappa}{\pi l_{\rm B}} \left(1+ \frac{3 \,\cal C}{4 \,\kappa}
\right).
\end{equation}
To dominant order, that is when ${\cal C}=0$, we recover as expected the planar 
result of Eq. \eqref{eq:sigmaeffsatplane}.
This suggests that to dominant plus sub-dominant order,
the effective charge only depends on the curvature,
irrespective of further geometrical details \cite{rque1}. We shall see
that this ``universality'' is broken by higher order terms
in ${\cal C}^2$.

\renewcommand{\theequation}{4.\arabic{equation}}
\setcounter{equation}{0}

\section{General formalism for 1:1 electrolyte} \label{1:1}
The above section motivates us to search for the potential
$\phi(x)$ in the ansatz form (\ref{finalphi1}) which is nothing but
the redefinition of the potential in terms of a new function $f(x)$
with simpler expansion property, as we will see later.
Inserting the ansatz (\ref{finalphi1}) into the PB equation (\ref{generalPB}), 
we obtain after some simple algebra the following differential equation 
for the $f$-function:
\begin{eqnarray}
\left[ f''(x) + \frac{\alpha}{x} f'(x) - f(x) \right] 
\left[ 1 - f^2(x) \right] & & \nonumber \\
- 2 f(x) \left[ f(x) + f'(x) \right] \left[ f(x) - f'(x) \right]
& = & 0 . \label{fequations}
\end{eqnarray}

With regard to the results of the preceding section, we expect that
the $f$-function can be written as
\begin{equation} \label{fg}
f(x) = t g(x) , 
\end{equation}
where the $x$-dependent function $t$ is defined in (\ref{deft}) and
the $g$-function is expected to have the following large-$x$ expansion
\begin{equation} \label{grepr}
g(x) = 1 + \sum_{k=1}^{\infty} \frac{1}{x^k} g_k(t) .
\end{equation}
In the DH limit $t\to 0$, $f$ is going to 0 as well and from (\ref{finalphi1}) 
we can write that $\phi(x) \sim 4 f(x)$.
Using the original expansion (\ref{potentialt}), we identify 
\begin{equation} \label{bcforg}
g_k(0) = \frac{B_{0k}}{4} \qquad \mbox{for all $k=1,2,\ldots$.} 
\end{equation}

From the definition of $t$ in Eq. (\ref{deft}), we have that
\begin{equation}
t'(x) = - \left( 1 + \frac{\alpha}{2x} \right) t , \qquad
t''(x) = \left[ \left( 1 + \frac{\alpha}{2x}\right)^2 
+ \frac{\alpha}{2 x^2} \right] t . 
\end{equation}
Inserting the representation (\ref{fg}) into Eq. (\ref{fequations}) and
using these relations, we obtain the differential equation for 
the $g$-function: 
\begin{eqnarray}
\left[ g''(x) -2 g'(x) + \frac{\alpha(2-\alpha)}{4 x^2} g(x) \right] 
\left[ 1 - t^2 g^2(x) \right] & & \nonumber \\
- 2 t^2 g(x) \left[ g'(x) - \frac{\alpha}{2x} g(x) \right] 
\left[\left( 2 + \frac{\alpha}{2 x} \right) g(x)  - g'(x) \right]
& = & 0 . \nonumber \\ & & \label{gequations}
\end{eqnarray}
We emphasize that a prime here means the {\em total} derivative with respect 
to $x$, including the function $t(x)$.
Like for instance, within the representation (\ref{grepr}) we have
\begin{equation}
g'(x) = - \sum_{k=1}^{\infty} \frac{1}{x^{k+1}} \left[ k g_k(t) +
\frac{\alpha}{2} t g'_k(t) \right] - \sum_{k=1}^{\infty} \frac{1}{x^k} t g'_k(t)
\end{equation}
and so on.
The point is that the total derivative with respect to $x$ keeps 
the power-law expansion in $1/x$ where each term is multiplied by 
a function depending on $t$ only.
This two-scale method permits us to determine recursively 
the functions $\{ g_k(t) \}$. 
Setting to zero the coefficient to $1/x$, we obtain a differential
equation for $g_1(t)$, supplemented by the BC $g_1(0) = B_{01}/4$
deduced from (\ref{bcforg}).
Setting to zero the coefficient to $1/x^2$, we obtain a differential equation 
for $g_2(t)$ which involves also the known function $g_1(t)$, supplemented by 
the BC $g_2(0) = B_{02}/4$, etc. 

\subsection{Cylindrical geometry}
We start with the cylindrical geometry for which $\alpha=1$. 
The BC (\ref{bcforg}) implies
\begin{equation}
g_k(0) = \frac{1}{2^k} \frac{\Gamma(\frac{1}{2}+k)}{k! \Gamma(\frac{1}{2}-k)} 
\end{equation} 
for all $k=1,2,\ldots$.

Setting to zero the coefficient to $1/x$ in (\ref{gequations}), 
$g_1(t)$ must obey the differential equation
\begin{equation}
t(1-t^2) g_1''(t) + (3+t^2) g_1'(t) + 2 t = 0 .
\end{equation}
The general solution of this equation, obtained by using the `variation of
constants' method, reads as
\begin{equation}
g_1(t) = c_1 + c_2 t^2 - (1+4c_2) \left( \ln t + \frac{1}{4 t^2} \right) .
\end{equation}
The integration constant $c_2$ is determined by the regularity of
$g_1(t)$ as $t\to 0$ as follows $c_2=-1/4$.
The BC $g_1(0) = -1/8$ leads to $c_1=-1/8$.
The consequent
\begin{equation}
g_1(t) = - \frac{1}{8} \left( 1 + 2 t^2 \right)
\end{equation}
is in full agreement with the previous result (\ref{finalphi2}).

To find $g_2(t)$, we set to zero the coefficient to $1/x^2$ in 
(\ref{gequations}) which, together with the knowledge of $g_1(t)$, implies
\begin{equation}
t(1-t^2) g_2''(t) + (3+t^2) g_2'(t) + \frac{t}{4}(2t^4-6t^2-15) = 0 .
\end{equation}
The general solution of this differential equation is
\begin{equation}
g_2(t) = c_1 + c_2 t^2 + \frac{t^4}{16} + 
\left( \frac{15}{8}-4c_2 \right) \left( \ln t + \frac{1}{4 t^2} \right) .
\end{equation}
The regularity of $g_2(t)$ at $t=0$ implies $c_2=15/32$ and the BC
$g_2(0)=9/128$ fixes $c_1=9/128$.
Thus we arrive at
\begin{equation}
g_2(t) = \frac{3^2}{2^7} + \frac{3\times 5}{2^5} t^2 + \frac{1}{4^2} t^4 .
\end{equation}

In the same way, e.g. by using the symbolic language {\it Mathematica},
we get
\begin{eqnarray}
g_3(t) & = & - \frac{3\times 5^2}{2^{10}} - \frac{163\times 3}{2^9} t^2 
- \frac{29}{2^7} t^4 - \frac{1}{4^3} t^6 , \\
g_4(t) & = & \frac{3\times(5\times 7)^2}{2^{15}} 
+ \frac{3\times 53\times 59}{2^{12}} t^2 
+ \frac{3^4\times 17}{2^{11}} t^4 \nonumber \\ & & 
+ \frac{43}{2^9} t^6 + \frac{1}{4^4} t^8 ,
\end{eqnarray}
etc.
In general, $g_k(t)$ with $k=1,2,\ldots$ turns out to be 
a finite polynomial of the $k$th order in the variable $t^2$.
This special and convenient property of the $g_k$-functions is present exclusively
for the case of the cylindrical geometry and the symmetric $1:1$ electrolyte.

\begin{figure}[htb]
\begin{center}
\includegraphics[width=0.5\textwidth,clip]{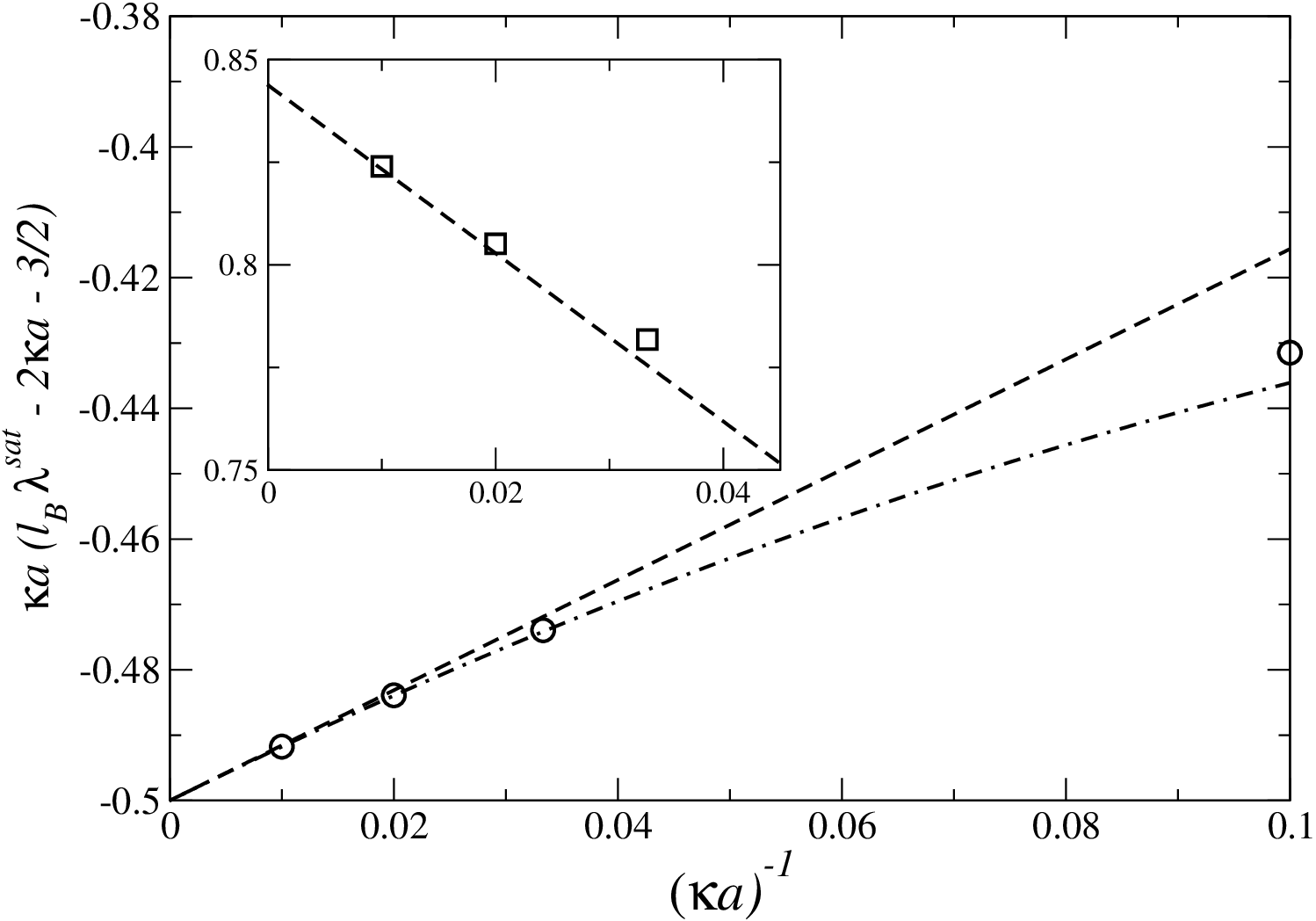}
\caption{Asymptotic expansion plots for cylinders in a 1:1 salt.
The main graph shows $\kappa a (\lambda_{\rm eff}^{\rm sat} l_B-2 \kappa a -3/2)$ as a function of $1/(\kappa a)$.
The dashed linear curve is for the first order correction predicted by Eq. \eqref{series}, $-1/2 + 27/(32 \kappa a)$,
while the parabolic dot-dashed curve is for $-1/2 + 27( \kappa a)^{-1}/32-131 (\kappa a)^{-2}/64$.
The symbols correspond to the saturated effective charge obtained from the numerical solution of Eq. \eqref{generalPB} with $\alpha=1$, and have been computed for $\kappa a=10$, 30, 50 and 100, as can be read on the $x$-axis. 
Inset : $(\kappa a)^2 [\lambda_{\rm eff}^{\rm sat} l_B-2 \kappa a -3/2-0.5/(\kappa a)]$ versus $1/(\kappa a)$. The linear 
dashed curve displays the prediction of Eq. \eqref{generalPB} for the quantity of interest, namely 
$27/32- 131(\kappa a)^{-1}/64$.}
\label{fig:cyl11} 
\end{center}
\end{figure}

Having $f(x)$ (\ref{fg}) with $g(x)$ (\ref{grepr}) truncated say at $k=4$,
the crucial function $A(\lambda)$ is again determined by the relation
(\ref{Alambda}).
In the saturation limit $\lambda\to\infty$, the requirement 
$f_{\rm sat}(\kappa a) = 1$ implies the iteratively generated large-$\kappa a$ 
expansion
\begin{eqnarray}
t_{\rm sat}(\kappa a) & = & 1 + \frac{3}{2^3} \frac{1}{\kappa a} -
\frac{5\times 7}{2^7} \frac{1}{(\kappa a)^2} + 
\frac{53\times 3^2}{2^{10}} \frac{1}{(\kappa a)^3} \nonumber \\ & &
-\frac{7369\times 5}{2^{15}} \frac{1}{(\kappa a)^4} + \cdots  
\end{eqnarray}
which goes beyond the previous one (\ref{tsat}).
Using the connection (\ref{prescr1}), the corresponding saturation value of 
the effective charge exhibits the large-$\kappa a$ expansion of the form
\begin{eqnarray}
\lambda_{\rm eff}^{\rm sat} l_{\rm B} & = & 2 \kappa a + \frac{3}{2} - 
\frac{1}{2} \frac{1}{\kappa a} + \frac{27}{32} \frac{1}{(\kappa a)^2}
\nonumber \\ & & - \frac{131}{64} \frac{1}{(\kappa a)^3} + 
O\left( \frac{1}{(\kappa a)^4} \right) . \label{series}
\end{eqnarray} 
Notice that in spite of the complicated large-$\kappa a$ expansion of 
$t_{\rm sat}(\kappa a)$, the corresponding expression for 
$\lambda_{\rm eff}^{\rm sat}$ is relatively simple.
The numerical checks of the coefficients to the $1/(\kappa a)$,
$1/(\kappa a)^2$, $\ldots$ terms are presented in Fig. \ref{fig:cyl11}.
Such a comparison poses the difficulty that the  
effective charge $\lambda_{\rm eff}$ be known with high precision
in the limit where the bare charge $\lambda$ diverges.
To this end, the Poisson-Boltzmann equation \eqref{generalPB} is solved numerically for a series 
of increasing bare charges, at a given value of $\kappa a$. The effective charge $\lambda_{\rm eff}$
is extracted from the far-field behaviour of the potential, that reads
\begin{equation}
\phi(r) = \frac{2\lambda_{\rm eff} l_{\rm B}}{\kappa a K_1(\kappa a)}
K_0(\kappa r) , \qquad r \to \infty. 
\end{equation}
A `finite-charge' scaling
analysis is subsequently performed: it is indeed straightforward to show that 
$\lambda_{\rm eff} - \lambda_{\rm eff}^{\rm sat} $ vanishes as $1/\lambda$,
when $\lambda\to \infty$. In practice, the above linear regime in $1/\lambda$ is well
reached whenever $\lambda> 10^6$. The saturated values thereby obtained are shown by the symbols
in all the graphs displayed. Once $\lambda_{\rm eff}^{\rm sat}$ is known, inspecting its
behaviour as a function of $\kappa a$, as performed in Fig. \ref{fig:cyl11}, allows form a stringent
test of the analytical prediction. The main graph reveals that beyond the dominant 
behaviour in $2 \kappa a +3/2$, the next correction to $\lambda_{\rm eff}^{\rm sat}$
is $-1/(2\kappa a)$, since the data shown extrapolate to $-1/2$ in the limit 
$1/(\kappa a)\to 0$. Besides, the next term predicted with prefactor $27/32$ 
brings significant improvement at large although finite $\kappa a$ (see the linear 
dashed line in the main graph). Yet, some (negative) curvature can be inferred from the
symbols shown and indeed, including the next term with prefactor $-131/64$ further enhances the
agreement. The inset offers a direct proof that expression \eqref{series}, with all terms, 
is a very plausible expansion for the saturated effective charge. Note that the quadratic
dashed-dotted curve in the main graph and the dashed line of the inset convey the same information,
in a different visual setting.

We would like to emphasize that our method of generating the large-$\kappa a$
expansion is technically very simple and we can generate within few seconds 
by using {\it Mathematica} also the next higher-order terms of the series 
(\ref{series}).

\subsection{Spherical geometry}
For the spherical $\alpha=2$ geometry, the BC (\ref{bcforg}) implies
\begin{equation}
g_k(0) = 0 
\end{equation}
for all $k=1,2,\ldots$.

Setting to zero the coefficient to $1/x$ in (\ref{gequations}), 
$g_1(t)$ fulfills the differential equation
\begin{equation}
t(1-t^2) g_1''(t) + (3+t^2) g_1'(t) + 4 t = 0 .
\end{equation}
The general solution of this equation reads
\begin{equation}
g_1(t) = c_1 + c_2 t^2 - 2(1+2c_2) \left( \ln t + \frac{1}{4 t^2} \right) .
\end{equation}
The regularity of $g_1(t)$ as $t\to 0$ fixes $c_2=-1/2$ and the BC 
$g_1(0)=0$ implies $c_1=0$.
Thus we have
\begin{equation}
g_1(t) = - \frac{1}{2} t^2 
\end{equation}
which agrees with the result (\ref{f}).

Setting to zero the coefficient to $1/x^2$ in (\ref{gequations}), 
we obtain the differential equation for $g_2(t)$:
\begin{equation}
t(1-t^2) g_2''(t) + (3+t^2) g_2'(t) + t(2t^4-7t^2-7) = 0 .
\end{equation}
The requirement of regularity as $t\to 0$ and the BC $g_2(0)=0$ 
imply the solution
\begin{equation}
g_2(t) = \frac{7}{8} t^2 + \frac{7}{24} t^4 + 
\frac{1}{2} \sum_{n=3}^{\infty} \frac{t^{2n}}{n^2(n^2-1)} .
\end{equation}
The function $g_2(t)$ is an infinite polynomial in $t^2$ with the convergence
radius $t\le 1$.
In particular, 
\begin{equation} \label{g21}
g_2(1) = 2 - \frac{\pi^2}{12} .
\end{equation}

\begin{figure}[htb]
\begin{center}
\includegraphics[width=0.5\textwidth,clip]{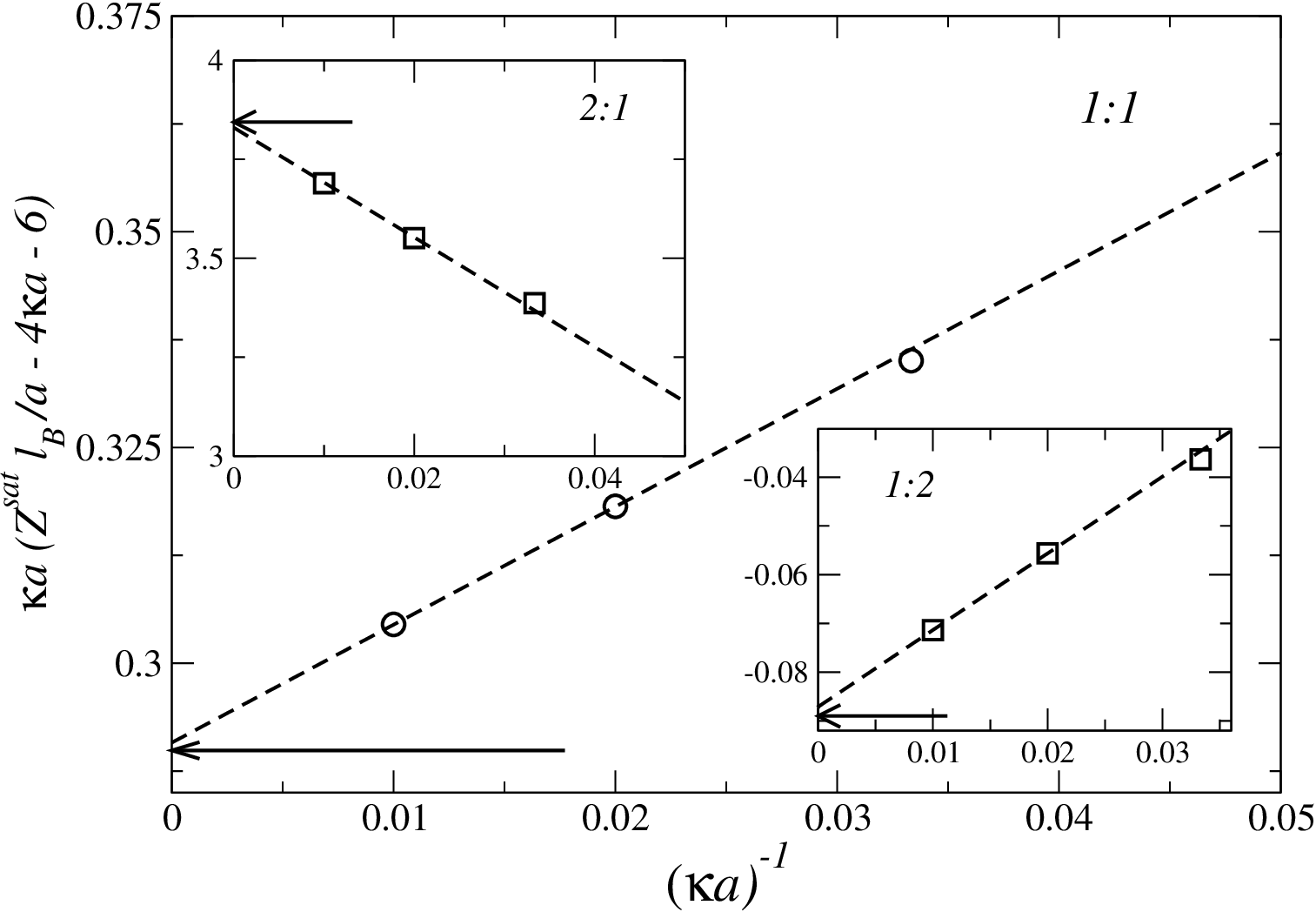}
\caption{Asymptotic expansion plots for charged spheres in a 1:1 salt (main graph),
together with asymmetric 1:2 and 2:1 salts (insets).
The main graph shows $\kappa a (Z_{\rm eff}^{\rm sat} l_B/a-4 \kappa a -6)$ as a function of $1/(\kappa a)$.
The dashed linear curve is for the line that passes through the two points associated to the largest values
of $\kappa a$ (50 and 100). The arrow is the predicted asymptotic limit $\pi^2/3-3$.
The symbols correspond to the saturated effective charge obtained from the numerical solution of Eq. \eqref{generalPB} with
$\alpha=2$.
The 2:1 inset shows the counterpart, $\kappa a (Z_{\rm eff}^{\rm sat} l_B/a-6 \kappa a -7)$
where the limit is predicted to be $\simeq 3.844$. The 1:2 inset is constructed similarly
from Eq. \eqref{eq:Sph12}, for which the arrow indicating the intercept is 
at -0.089.}
\label{fig:sph} 
\end{center}
\end{figure}

In the saturation limit $\lambda\to\infty$, the requirement 
$f_{\rm sat}(\kappa a) = 1$ implies the iteratively generated 
large-$\kappa a$ expansion
\begin{equation}
t_{\rm sat}(\kappa a) =  1 + \frac{1}{2} \frac{1}{\kappa a} - \left( 
\frac{5}{4} - \frac{\pi^2}{12} \right) \frac{1}{(\kappa a)^2} + \cdots .  
\end{equation}
Here, we used that $g_2(1)$ is given by (\ref{g21}).
We cannot go beyond the indicated order because the next term needs
the iteration with the diverging value of $g_2(1+1/(2\kappa a))$.
This indicates that the next-order singular term has a form different
from $1/(\kappa a)^3$. 

Using the prescription (\ref{prescr2}), the corresponding saturation value of 
the effective charge exhibits the large-$\kappa a$ expansion of the form
\begin{equation}
Z_{\rm eff}^{\rm sat} \frac{l_{\rm B}}{a} = 4 \kappa a + 6 + 
\left( \frac{\pi^2}{3} - 3 \right) \frac{1}{\kappa a} + 
o\left( \frac{1}{\kappa a} \right) .
\end{equation} 
The numerical check of the prefactor to the third $1/(\kappa a)$ term 
is presented in Fig. \ref{fig:sph}. Indirectly, the plot also assesses 
the correctness of the dominant terms $4\kappa a +6$.

\renewcommand{\theequation}{5.\arabic{equation}}
\setcounter{equation}{0}

\section{2:1 electrolyte} \label{2:1}
For the asymmetric $2:1$ electrolyte in contact with the cylindrical
$\alpha=1$ or spherical $\alpha=2$ colloids, the PB equation 
in the reduced distance $x=\kappa r$ takes the form
\begin{equation} \label{PB21}
\phi''(x) + \frac{\alpha}{x} \phi'(x) = \frac{1}{3} \left[
{\rm e}^{\phi(x)} - {\rm e}^{-2\phi(x)} \right] .
\end{equation}

Being motivated by the exact planar solution (\ref{Gouy}), we search
the electrostatic potential in an ansatz form
\begin{equation} \label{ansatz21}
\phi(x) = \ln \left\{ 1 + \frac{6f(x)}{\left[ 1-f(x) \right]^2} \right\} .
\end{equation}
Introducing
\begin{equation} 
t(x) = \frac{A(\lambda)}{6} \frac{{\rm e}^{-x}}{x^{\alpha/2}} ,
\end{equation}
$f(x)\sim t(x)$ is the leading large-distance form.
Inserting the ansatz (\ref{ansatz21}) into the PB equation (\ref{PB21}),
we obtain the differential equation for the $f$-function: 
\begin{widetext}
\begin{eqnarray} 
\left[ f''(x) + \frac{\alpha}{x} f'(x) - f(x) \right] 
\left[ 1 - f^2(x) \right] \left[ f^2(x) + 4 f(x) + 1 \right] & & \nonumber \\
- 2 \left[ f^3(x)+3f^2(x)+3f(x)-1 \right] 
\left[ f(x) + f'(x) \right] \left[ f(x) - f'(x) \right] & = & 0 . 
\label{f21equations}
\end{eqnarray}
We again expect that the $f$-function is expressible as a series
\begin{equation}
f(x) = t g(x) , \qquad g(x) = 1 + \sum_{k=1}^{\infty} \frac{1}{x^k} g_k(t) .
\end{equation}
The $g$-function satisfies the differential equation
\begin{eqnarray}
\left[ g''(x) -2 g'(x) + \frac{\alpha(2-\alpha)}{4 x^2} g(x) \right] 
\left[ 1 - t^2 g^2(x) \right] \left[ t^2 g^2(x) + 4 t g(x) +1 \right] 
& & \nonumber \\
- 2 t \left[ t^3 g^3(x) + 3 t^2 g^2(x) + 3 t g(x) - 1 \right] 
\left[ g'(x) - \frac{\alpha}{2x} g(x) \right] 
\left[\left( 2 + \frac{\alpha}{2 x} \right) g(x)  - g'(x) \right]
& = & 0 . \label{gequations21}
\end{eqnarray}
\end{widetext}

\subsection{Cylindrical geometry}
For the cylindrical $\alpha=1$ geometry, setting to zero the coefficient to 
$1/x$ in (\ref{gequations21}), $g_1(t)$ obeys the differential equation
\begin{eqnarray} 
t(1+4t-4t^3-t^4) g_1''(t) & & \nonumber \\
+ (3+8t+12t^2+t^4) g_1'(t) & & \nonumber \\
+ 2(-1+3t+3t^2+t^3) & = & 0 . \label{eqn}
\end{eqnarray}
The solution of this equation with the BC $g_1(0)=-1/8$, regular as $t\to 0$, 
is found to be
\begin{equation}
g_1(t) = \frac{29}{24} - \frac{2}{3} t - \frac{1}{12} t^2 - 
\frac{4}{3} \frac{1}{1+t} .
\end{equation}
Similarly, we obtain
\begin{eqnarray}
g_2(t) & = & - \frac{2543}{2^7\times 3^2} + \frac{11}{2\times 3^2} t 
+ \frac{5}{2^5} t^2 +\frac{1}{3^2} t^3 + \frac{1}{2^4\times3^2} t^4 
\nonumber \\ & & + \frac{2\times 5}{3} \frac{1}{1+t} 
- \frac{5\times 7}{2\times 3^2} \frac{1}{(1+t)^2}
+ \frac{2^3}{3^2} \frac{1}{(1+t)^3} , \nonumber \\ & & \\
g_3(t) & = & \frac{5\times 41\times 659}{2^{10}\times 3^3} 
+ \frac{7^2}{2^4\times 3^2} t - \frac{617}{2^9\times 3} t^2 
-\frac{17}{2^2\times 3^3} t^3 \nonumber \\ & &
- \frac{13\times 19}{2^7\times 3^3} t^4 - \frac{1}{2^3\times 3^2} t^5 
- \frac{1}{2^6\times 3^3} t^6 \nonumber \\ & & 
- \frac{193}{2\times 3^2} \frac{1}{1+t} 
+ \frac{11173}{2^5\times 3^3} \frac{1}{(1+t)^2} \nonumber \\ & &
- \frac{4649}{2^4\times 3^3} \frac{1}{(1+t)^3} 
+ \frac{43}{3^2} \frac{1}{(1+t)^4} - \frac{2^5}{3^3} \frac{1}{(1+t)^5} ,
\nonumber \\ & &
\end{eqnarray}
etc.
In general, $g_k(t)$ $(k=1,2,\ldots)$ is the sum of two finite polynomial, 
one of the $2k$th order in $t$ and the other of the $(2k-1)$th order in 
$1/(1+t)$.

\begin{figure}[ht]
\begin{center}
\includegraphics[width=0.5\textwidth,clip]{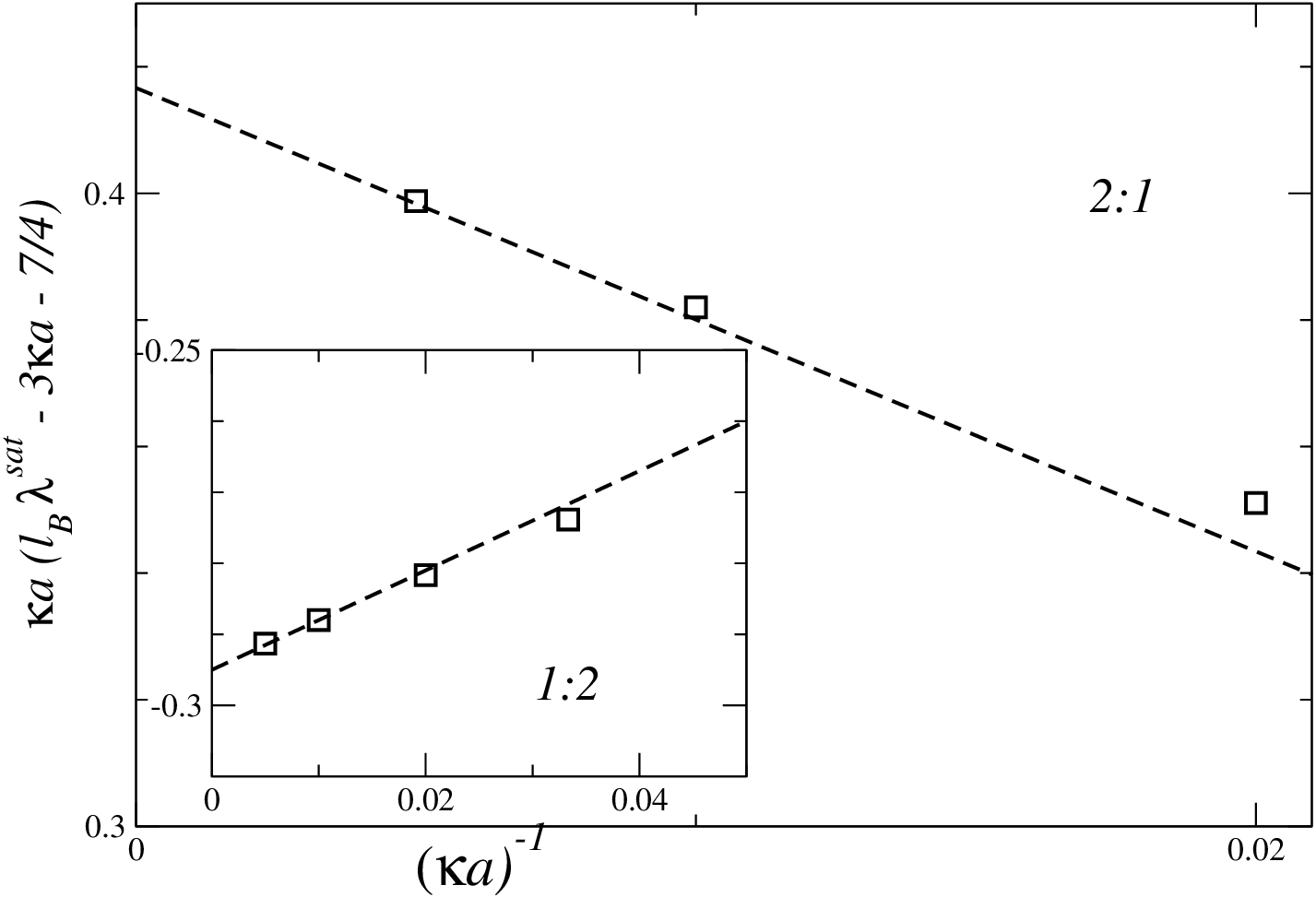}
\caption{Asymptotic expansion plots for cylinders in  2:1 and 1:2 salts.
The main graph for the 2:1 case shows  $\kappa a (\lambda_{\rm eff}^{\rm sat} l_B-3 \kappa a -7/4)$ versus $1/(\kappa a)$.
The linear 
dashed curve displays the prediction embodied in Eq. \eqref{eq:lasat21}, $5/12-703(\kappa a)^{-1}/192$.
The inset displays expression 
\eqref{eq:forgraph} as a function of $1/(\kappa a)$.
The dashed linear curve is for the prediction of Eq. \eqref{eq:lasat12}, $-0.295 + 0.700/( \kappa a)$. 
As in previous figures, the symbols show the data obtained from solving numerically the
Poisson-Boltzmann equation.}
\label{fig:cyl12et21} 
\end{center}
\end{figure}

In the saturation limit $\lambda\to\infty$, the requirement 
$f_{\rm sat}(\kappa a) = 1$ implies the iteratively generated large-$\kappa a$ 
expansion
\begin{eqnarray}
t_{\rm sat}(\kappa a) & = & 1 + \frac{5}{2^3\times 3} \frac{1}{\kappa a} +
\frac{5\times 41}{2^7\times 3^2} \frac{1}{(\kappa a)^2} \nonumber \\
& & - \frac{12583}{2^{10}\times 3^2} \frac{1}{(\kappa a)^3} + \cdots 
\equiv \frac{A_{\rm sat}}{6} \frac{{\rm e}^{-\kappa a}}{\sqrt{\kappa a}} .  
\phantom{aaa}
\end{eqnarray}
Using the prescription (\ref{prescr1}), for large values of $\kappa a$ 
the saturation value of the effective charge behaves as
\begin{equation}
\lambda_{\rm eff}^{\rm sat} l_{\rm B} = 3 \kappa a + \frac{7}{4} + 
\frac{5}{12} \frac{1}{\kappa a} - \frac{703}{192} \frac{1}{(\kappa a)^2}
+ O\left( \frac{1}{(\kappa a)^3} \right) .
\label{eq:lasat21}
\end{equation} 
The first two terms of this series have been obtained in Ref. \cite{Tellez04}.
The numerical checks of the coefficients to the $1/(\kappa a)$ and
$1/(\kappa a)^2$ terms are presented in the main graph of Fig. \ref{fig:cyl12et21}.

\subsection{Spherical geometry}
For the spherical $\alpha=2$ geometry, setting to zero the coefficient to 
$1/x$ in (\ref{gequations21}), $g_1(t)$ fulfills the equation 
\begin{eqnarray} 
t(1+4t-4t^3-t^4) g_1''(t) & & \nonumber \\
+ (3+8t+12t^2+t^4) g_1'(t) & & \nonumber \\
+ 4(-1+3t+3t^2+t^3) & = & 0 . 
\end{eqnarray}
Considering the BC $g_1(0)=0$, the solution is
\begin{equation}
g_1(t) = \frac{8}{3} - \frac{4}{3} t - \frac{1}{6} t^2
- \frac{8}{3} \frac{1}{1+t} . 
\end{equation}

Setting to zero the coefficient to $1/x^2$ in (\ref{gequations21}), 
we obtain the differential equation for $g_2(t)$ of the form
\begin{equation} \label{diffeq}
P(t) g_2''(t) + Q(t) g_2'(t) + R(t) = 0 , 
\end{equation}
where
\begin{eqnarray}
P(t) & = & 9 t (1+t)^5 (1+3t-3t^2-t^3) , \\
Q(t) & = & 9 (1+t)^4 (3+8t+12t^2+t^4) , \\
R(t) & = & 78 + 51 t + 84 t^2 + 252 t^3 -1912 t^4 \nonumber \\
& & - 2560 t^5 -1860 t^6 - 430 t^7 + 314 t^8 \nonumber \\
& & + 189 t^9 + 32 t^{10} + 2 t^{11} .
\end{eqnarray}
The function $g_2(t)$ is an infinite polynomial in $t$ which
diverges for all $t>1$.
The value of $g_2$ at $t=1$ is of our primary interest. 
Denoting $g_2(1)\equiv {\cal A}$, by using {\it Mathematica} 
it can be shown that
\begin{eqnarray}
{\cal A} & = & - \frac{4}{9} - \frac{\pi^2}{6} 
- \frac{2}{\sqrt{3}} \ln(2-\sqrt{3}) + \frac{1}{3} [\ln(2-\sqrt{3})]^2 
\nonumber \\ & & - \frac{1}{6} \left[ \ln(3+\sqrt{3}) \right]^2 
- \frac{1}{3} \ln(2-\sqrt{3}) \ln(3-\sqrt{3}) \nonumber \\ & &
-\frac{1}{3} {\rm Li}_2\left(\frac{1}{6}(3-\sqrt{3})\right)  
-\frac{1}{3} {\rm Li}_2(-2+\sqrt{3}) \nonumber \\  & \simeq & -0.27962 ,
\end{eqnarray}
where ${\rm Li}_2(z)$ is the polylogarithm function defined by
\begin{equation}
{\rm Li}_2(z) = \sum_{k=1}^{\infty} \frac{z^k}{k^2} , \qquad \vert z\vert \le 1 .
\end{equation}
  
The saturation condition $f_{\rm sat}(\kappa a) = 1$ implies 
the large-$\kappa a$ expansion
\begin{eqnarray}
t_{\rm sat}(\kappa a) & = & 1 + \frac{1}{6} \frac{1}{\kappa a} 
+ \left( \frac{7}{36} - {\cal A} \right) \frac{1}{(\kappa a)^2} \cdots
\nonumber \\ & \equiv & 
\frac{A_{\rm sat}}{6} \frac{{\rm e}^{-\kappa a}}{\kappa a} .  
\end{eqnarray}
With respect to the prescription (\ref{prescr2}), the saturation value of 
the effective charge behaves for large values of $\kappa a$ as follows
\begin{eqnarray}
Z_{\rm eff}^{\rm sat} \frac{l_{\rm B}}{a} & = & 6 \kappa a + 7 + 
\left( \frac{13}{6} - 6 {\cal A} \right) \frac{1}{\kappa a} 
+ o\left( \frac{1}{\kappa a} \right) 
\nonumber \\ & \simeq & 6 \kappa a + 7 + 3.844 \frac{1}{\kappa a} 
+ o\left( \frac{1}{\kappa a} \right)  .
\end{eqnarray} 
The first two terms of this expansion are in full agreement with 
the result of Ref. \cite{Tellez04}.
The prefactor to the $1/(\kappa a)$ term is checked against the numerical
resolution in Fig. \ref{fig:sph}.

\renewcommand{\theequation}{6.\arabic{equation}}
\setcounter{equation}{0}

\section{1:2 electrolyte} \label{1:2}
For the asymmetric $1:2$ electrolyte, the PB equation takes the form
\begin{equation} \label{PB12}
\phi''(x) + \frac{\alpha}{x} \phi'(x) = \frac{1}{3} \left[
{\rm e}^{2\phi(x)} - {\rm e}^{-\phi(x)} \right] .
\end{equation}
With regard to the exact planar solution (\ref{Gouy2}), the electrostatic 
potential is searched in an ansatz form
\begin{equation} \label{ansatz12}
\phi(x) = - \ln \left\{ 1 - \frac{6f(x)}{\left[ 1+f(x) \right]^2} \right\} .
\end{equation}
Introducing
\begin{equation} 
t(x) = \frac{A(\lambda)}{6} \frac{{\rm e}^{-x}}{x^{\alpha/2}} ,
\end{equation}
$f(x)\sim t(x)$ in the leading large-distance order.

Inserting (\ref{ansatz12}) into the PB equation (\ref{PB12}),
the $f$-function obeys the differential equation 
\begin{widetext}
\begin{eqnarray} 
\left[ f''(x) + \frac{\alpha}{x} f'(x) - f(x) \right] 
\left[ 1 - f^2(x) \right] \left[ f^2(x) - 4 f(x) + 1 \right] & & \nonumber \\
- 2 \left[ f^3(x)-3f^2(x)+3f(x)+1 \right] 
\left[ f(x) + f'(x) \right] \left[ f(x) - f'(x) \right] & = & 0 . 
\label{f12equations}
\end{eqnarray}
Writing the $f$-function as a series
\begin{equation}
f(x) = t g(x) , \qquad g(x) = 1 + \sum_{k=1}^{\infty} \frac{1}{x^k} g_k(t) ,
\end{equation}
the $g$-function satisfies the differential equation
\begin{eqnarray}
\left[ g''(x) -2 g'(x) + \frac{\alpha(2-\alpha)}{4 x^2} g(x) \right] 
\left[ 1 - t^2 g^2(x) \right] \left[ t^2 g^2(x) - 4 t g(x) +1 \right] 
& & \nonumber \\
- 2 t \left[ t^3 g^3(x) - 3 t^2 g^2(x) + 3 t g(x) + 1 \right] 
\left[ g'(x) - \frac{\alpha}{2x} g(x) \right] 
\left[\left( 2 + \frac{\alpha}{2 x} \right) g(x)  - g'(x) \right]
& = & 0 . \label{gequations12}
\end{eqnarray}
\end{widetext}

\subsection{Cylindrical geometry}
For $\alpha=1$, setting to zero the coefficient to $1/x$ in 
(\ref{gequations12}), $g_1(t)$ must obey the equation
\begin{eqnarray} 
t(1-4t+4t^3-t^4) g_1''(t) & & \nonumber \\
+ (3-8t+12t^2+t^4) g_1'(t) & & \nonumber \\
+ 2(1+3t-3t^2+t^3) & = & 0 . \nonumber \\ & & \label{eqn2}
\end{eqnarray}
The regular solution of this equation with the BC $g_1(0)=-1/8$ reads
\begin{equation}
g_1(t) = \frac{29}{24} + \frac{2}{3} t - \frac{1}{12} t^2 - 
\frac{4}{3} \frac{1}{1-t} .
\end{equation}
In the same way, we get
\begin{eqnarray}
g_2(t) & = & - \frac{2543}{2^7\times 3^2} - \frac{11}{2\times 3^2} t 
+ \frac{5}{2^5} t^2 -\frac{1}{3^2} t^3 + \frac{1}{2^4\times 3^2} t^4 
\nonumber \\ & & + \frac{2\times 5}{3} \frac{1}{1-t} 
- \frac{5\times 7}{2\times 3^2} \frac{1}{(1-t)^2}
+ \frac{2^3}{3^2} \frac{1}{(1-t)^3} , \nonumber \\ & & \\
g_3(t) & = & \frac{5\times 41\times 659}{2^{10}\times 3^3} 
- \frac{7^2}{2^4\times 3^2} t - \frac{617}{2^9\times 3} t^2 
+ \frac{17}{2^2\times 3^3} t^3 \nonumber \\ & &
- \frac{13\times 19}{2^7\times 3^3} t^4 + \frac{1}{2^3\times 3^2} t^5 
- \frac{1}{2^6\times 3^3} t^6 \nonumber \\ & & 
- \frac{193}{2\times 3^2} \frac{1}{1-t} 
+ \frac{11173}{2^5\times 3^3} \frac{1}{(1-t)^2} \nonumber \\ & &
- \frac{4649}{2^4\times 3^3} \frac{1}{(1-t)^3} 
+ \frac{43}{3^2} \frac{1}{(1-t)^4} - \frac{2^5}{3^3} \frac{1}{(1-t)^5} ,
\nonumber \\ & &
\end{eqnarray}
etc.

In the saturation limit $\lambda\to\infty$, the requirement 
$f_{\rm sat}(\kappa a) = 2-\sqrt{3}$ implies the large-$\kappa a$ 
expansion
\begin{eqnarray}
t_{\rm sat}(\kappa a) & = & 2-\sqrt{3} + \left( -\frac{67}{12} + 
\frac{79}{8\sqrt{3}} \right) \frac{1}{\kappa a} \nonumber \\ & &
+ \left( \frac{17053}{576} - \frac{19765}{384\sqrt{3}} \right) 
\frac{1}{(\kappa a)^2} \nonumber \\ & &
+ \left( - \frac{751423}{4608} + \frac{869059}{3072\sqrt{3}} \right) 
\frac{1}{(\kappa a)^3} + \cdots \nonumber \\ & \equiv & 
\frac{A_{\rm sat}}{6} \frac{{\rm e}^{-\kappa a}}{\sqrt{\kappa a}} .  
\phantom{aaa}
\end{eqnarray}
With the aid of (\ref{prescr1}), for large values of $\kappa a$ 
the saturation value of the effective charge behaves as
\begin{eqnarray}
\lambda_{\rm eff}^{\rm sat} l_{\rm B} & = & 3 (2-\sqrt{3}) \kappa a + 
\left( -\frac{29}{2} + \frac{35\sqrt{3}}{4} \right) \nonumber \\ & & 
+ \left( \frac{491}{6} - \frac{569}{4\sqrt{3}} \right) 
\frac{1}{\kappa a} \nonumber \\ & &
+ \left( -\frac{43519}{96} + \frac{50329}{64\sqrt{3}} \right) 
\frac{1}{(\kappa a)^2} + O\left( \frac{1}{(\kappa a)^3} \right)   
\nonumber \\ & \simeq &
0.804 \,\kappa a + 0.655 - \frac{0.295}{\kappa a} \nonumber \\ & & 
+ \frac{0.700}{(\kappa a)^2} + O\left( \frac{1}{(\kappa a)^3} \right) .
\label{eq:lasat12}
\end{eqnarray} 
The first two terms of this series have been obtained in Ref. \cite{Tellez04}.
The numerical checks of the coefficients to the $1/(\kappa a)$ and 
$1/(\kappa a)^2$ terms are presented in Fig. \ref{fig:cyl12et21}.
The quantity plotted in the inset is thus 
\begin{equation}
\kappa a\left[\lambda_{\rm eff}^{\rm sat} l_{\rm B} - 3 (2-\sqrt{3}) \kappa a + 
\left( \frac{29}{2} - \frac{35\sqrt{3}}{4} \right) \right] .
\label{eq:forgraph}
\end{equation}

\subsection{Spherical geometry}
For $\alpha=2$, setting to zero the coefficient to $1/x$ in 
(\ref{gequations12}), $g_1(t)$ obeys the equation
\begin{eqnarray} 
t(1-4t+4t^3-t^4) g_1''(t) & & \nonumber \\
+ (3-8t+12t^2+t^4) g_1'(t) & & \nonumber \\
+ 4(1+3t-3t^2+t^3) & = & 0 . 
\end{eqnarray}
The regular solution of this equation with the BC $g_1(0)=0$ reads
\begin{equation}
g_1(t) = \frac{8}{3} + \frac{4}{3} t - \frac{1}{6} t^2 - 
\frac{8}{3} \frac{1}{1-t} .
\end{equation}

Setting to zero the coefficient to $1/x^2$ in (\ref{gequations12}), 
we obtain the differential equation for $g_2(t)$ of type (\ref{diffeq}) 
with the polynomial coefficients
\begin{eqnarray}
P(t) & = & 9 t (1-t)^5 (1-3t-3t^2-t^3) , \\
Q(t) & = & 9 (1-t)^4 (3-8t+12t^2+t^4) , \\
R(t) & = & -78 + 51 t - 84 t^2 + 252 t^3 + 1912 t^4 \nonumber \\
& & - 2560 t^5 + 1860 t^6 - 430 t^7 - 314 t^8 \nonumber \\
& & + 189 t^9 - 32 t^{10} + 2 t^{11} .
\end{eqnarray}
For our purpose, the value of $g_2$ at $t=2-\sqrt{3}$ will be important. 
Denoting $g_2(2-\sqrt{3})\equiv {\cal B}$, using {\it Mathematica} we got
\begin{eqnarray}
{\cal B} & = & -\frac{130}{9} + 10\sqrt{3} - \frac{\pi^2}{9} 
- \frac{4}{\sqrt{3}} \ln 2 - 2 \sqrt{3} \ln 3 \nonumber \\ & &
+ \frac{1}{3} \left[ \ln 12 - \frac{7}{\sqrt{3}} \right] \ln(2-\sqrt{3})
+ \frac{2}{3} [\ln(2-\sqrt{3})]^2 \nonumber \\ & & 
- \frac{29}{3\sqrt{3}} \ln(\sqrt{3}-1) + \frac{5}{3\sqrt{3}} \ln(\sqrt{3}+1) 
\nonumber \\ & & - \frac{4}{3} {\rm Li}_2(-2+\sqrt{3}) 
+ \frac{1}{3} {\rm Li}_2(-6+4\sqrt{3}) \nonumber \\  & \simeq & 1.71475 .
\end{eqnarray}
  
In the saturation regime, the condition $f_{\rm sat}(\kappa a) = 2-\sqrt{3}$ 
implies the large-$\kappa a$ expansion
\begin{eqnarray}
t_{\rm sat}(\kappa a) & = & (2-\sqrt{3}) + \left( -\frac{35}{3} + 
\frac{41}{2\sqrt{3}} \right) \frac{1}{\kappa a} \nonumber \\ & &
+ \left[ \frac{1123}{18} - \frac{1291}{12\sqrt{3}} - (2-\sqrt{3}) {\cal B}
\right] \frac{1}{(\kappa a)^2} \nonumber \\ & & + \cdots
\equiv \frac{A_{\rm sat}}{6} \frac{{\rm e}^{-\kappa a}}{\kappa a} .
\end{eqnarray}
Based on (\ref{prescr2}), the large-$\kappa a$ expansion of the saturation 
value of the effective charge is obtained in the form
\begin{eqnarray}
Z_{\rm eff}^{\rm sat} \frac{l_{\rm B}}{a} & = &  6 (2-\sqrt{3}) \kappa a + 
\left( -58 + 35\sqrt{3} \right) \nonumber \\ & &
+ \left[ \frac{913}{3} - \frac{1045}{2\sqrt{3}} - 6(2-\sqrt{3}) {\cal B} 
\right] \frac{1}{\kappa a} + o\left( \frac{1}{\kappa a} \right) \nonumber 
\\ & \simeq &  1.608 \kappa a + 2.622 - 0.089 \frac{1}{\kappa a} + 
o\left( \frac{1}{\kappa a} \right) . 
\label{eq:Sph12}
\end{eqnarray} 
The first two terms of the expansion coincide with those obtained in 
Ref. \cite{Tellez04}.
The prefactor to the third $1/(\kappa a)$ term is checked against
numerics in Fig.  \ref{fig:sph}.

\renewcommand{\theequation}{7.\arabic{equation}}
\setcounter{equation}{0}

\section{Conclusion} \label{Conclusion}
In this work, we have revisited the analytical results following from a multiple scale expansion 
of the non-linear Poisson-Boltzmann equation, for both cylindrical and spherical macro-ions. 
The corresponding planar case is analytically solvable. Three types of electrolyte 
have been addressed: symmetric ones where the co-and counter-ions bear the same charge 
in absolute value (1:1 case), as well as asymmetric 1:2 and 2:1 situations. The latter two cases are not 
equivalent due to the non-linear nature of the differential equation to be solved,
although they can yield the same Debye length.
Inspecting the structure of the double series appearing intimates that 
a partial resummation can be performed. In doing so, and restricting to the 1:1 case for the sake of simplicity,
the dimensionless electrostatic potential $\phi$
appears to depend on radial distance $r$ through 
\begin{equation}
\phi(r) \, =\, 2\, \ln \left( \frac{1+f(x)}{1-f(x)} \right) 
\label{eq:7.1}
\end{equation}
where $x=\kappa r$, and
\begin{equation}
f(x) \,=\, t \left[ 1+\sum_{k=1}^\infty \frac{g_k(t)}{x^k}
\right] \quad;\quad
t \,=\, \frac{A}{4} \frac{{\rm e}^{-x}}{x^{\alpha/2}} .
\label{eq:generic}
\end{equation}
Here, $\alpha$ is a fingerprint of geometry (more precisely, of curvature, with 
$\alpha=0$ for plates, $\alpha=1$ for cylinders, $\alpha=2$ for spheres)
and $A$ parameterizes the solution: different values of $A$ correspond to different bare charges.
The saturation phenomenon means that while $A$ changes in some finite interval $[0,A_{\text{sat}}]$,
the bare charge varies between 0 and $\infty$. More precisely, 
since one has $\phi \sim t$ for $r\to \infty$,
$A$ is directly related to the effective charge of the macro-ion.
For 1:2 and 2:1 electrolytes, relation \eqref{eq:7.1} changes to some extent
(see Eqs. \eqref{ansatz21} and \eqref{ansatz12}),
while the relation between $f$, $t$ and $A$ in Eq.
\eqref{eq:generic} is essentially unaffected.

The planar case is such that $\kappa a \to \infty$, with $f(x) = t = A \,e^{-x}/4$. 
As a consequence, our family of solutions is of ``quasi-planar'' type, which is of course quite expected 
in the limit where the macro-ion radius $a$ is much larger than the Debye length $1/\kappa$. Yet, the details 
of this quasi-planarity are non trivial, and are such that particularly convenient expansion 
properties ensue in the asymptotic limit $\kappa a \to \infty$.
As an illustration, we have computed saturated effective charges (meaning in the limit where the
macro-ion bare charge becomes very large) where our scheme yields an exact expansion 
in inverse powers of $\kappa a$. Indeed, a careful numerical calculation of the same quantities from 
solving directly the Poisson-Boltzmann equation, allows to check, term by term, the predicted 
expansion. This requires an extrapolation procedure, which has been presented, for extracting the saturation 
values from results that are necessarily obtained at finite although large bare charges.

So far, not enough is known on the planar case for different asymmetries than 1:2 and 2:1,
so that our approach cannot be generalized to such situations. What misses is the explicit structure 
of the counterpart of Eqs. \eqref{eq:7.1}, \eqref{ansatz21} and \eqref{ansatz12} in these cases \cite{Tellez11}.

\begin{acknowledgments}
L. \v{S}. is grateful to LPTMS for its hospitality.
The support received from the grant VEGA No. 2/0015/15 is acknowledged.
\end{acknowledgments}


\begin{thebibliography}{10}
\bibitem{rque50}
The mean-field treatment adopted does not discriminate the potential created by a disk in 2D from 
that of an infinite cylinder in 3D. The treatment exhibits in this sense some geometrical 
degeneracy.
\bibitem{rque1}
It can be checked from the expressions given in \cite{Aubouy03} --and again considering only the dominant plus first sub-dominant
contributions-- that beyond the saturation limit, a similar remark holds for the
full functional dependence of effective charges as a function of bare charges, when expressed
in terms of colloid curvature. Of course, this feature is not restricted to 1:1 electrolytes.
\end{thebibliography}
\end{document}